\documentstyle[psfig,12pt]{llncs} 
\def\VEC#1{{\mathchoice%
{\mbox{\boldmath$\displaystyle #1$}}%
{\mbox{\boldmath$\textstyle #1$}}%
{\mbox{\boldmath$\scriptstyle #1$}}%
{\mbox{\boldmath$\scriptscriptstyle #1$}}}}

\newcommand{\OL}{\overline}

\newcommand{\TR}{{\rm tr}}

\newcommand{\FUN}[1]{{\rm #1}}
\newcommand{\MAT}[1]{{\bf #1}}
\newcommand{\LAB}[1]{{\sf #1}}
\newcommand{\EMB}[1]{{\VEC #1}}
\newcommand{\FRAC}[2]{\mbox{$\frac{#1}{#2}$}}
\newcommand{\SQRT}[1]{\mbox{$\sqrt{#1}$}}
\newcommand{\KET}[1]{|\,#1\,\rangle}
\newcommand{\BRA}[1]{\langle\,#1\,|}
\newcommand{\AVG}[1]{\langle\,#1\,\rangle}
\newcommand{\HALF}{{\FRAC{1}{2}}}
\newcommand{\RHO}{{\EMB\rho}}
\newcommand{\VRHO}{{\EMB\varrho}}
\newcommand{\EP}{\VEC{E}_{+}}
\newcommand{\EM}{\VEC{E}_{-}}

\newcommand{\EPM}{\VEC{E}_{\pm}}
\newcommand{\EMP}{\VEC{E}_{\mp}}

\newcommand{\IP}{\VEC{I}_{+}}
\newcommand{\IM}{\VEC{I}_{-}}
\newcommand{\IPM}{\VEC{I}_{\pm}}
\newcommand{\IMP}{\VEC{I}_{\mp}}
\newcommand{\II}{\VEC{1}}
\newcommand{\IX}{\VEC{I}_\LAB{x}}
\newcommand{\IY}{\VEC{I}_\LAB{y}}
\newcommand{\IZ}{\VEC{I}_\LAB{z}}

\renewcommand{\footnoterule}{\vspace{3pt}%
\kern-3pt\hrule width 2 true cm\kern 2.6pt}
\renewcommand{\arraystretch}{1.3}

\newcommand{\skiplinehalf}{\vspace{10pt}}

\begin{document}% comes before \maketitle in this style

\title{Principles and demonstrations of quantum
information processing by NMR spectroscopy\thanks{
Portions of this survey were presented at the AeroSense
Workshop on Photonic Quantum Computing II, held in Orlando,
Florida on April 16, 1998, at the Dagstuhl Seminar
on Quantum Algorithms, held in Schloss Dagstuhl,
Germany on May 10 -- 15, 1998, and at the Workshop
on Quantum Information, Decoherence and Chaos, held
on Heron Island, Australia September 21 -- 25, 1998;
this paper is an updated and extended version of one
published in the proceedings of the AeroSense meeting,
available as vol.\ 3385 from the International Society for
Optical Engineering, 1000 20th St., Bellingham, WA 98225, USA.}}

\author{T.~F.~Havel\inst{1}, S.~S.~Somaroo\inst{1},
C.-H.~Tseng\inst{2} and D.~G.~Cory\inst{3}}

\institute{BCMP, Harvard Medical School, Boston, MA 02115
\and Harvard-Smithsonian Center for Astrophysics, Cambridge, MA 02138
\and Nuclear Engineering, MIT, Cambridge, MA 02139}

%\titlerunning{Quantum information processing by NMR}
%\authorrunning{Havel, Somaroo, Tseng and Cory}

\maketitle

\begin{abstract}
This paper surveys our recent research on quantum information
processing by nuclear magnetic resonance (NMR) spectroscopy.
We begin with a geometric introduction to the
NMR of an ensemble of indistinguishable spins,
and then show how this geometric interpretation is
contained within an algebra of multispin product operators.
This algebra is used throughout the rest of the paper
to demonstrate that it provides a facile framework within
which to study quantum information processing more generally.
The implementation of quantum algorithms by NMR depends
upon the availability of special kinds of mixed states,
called pseudo-pure states, and we consider a
number of different methods for preparing these states,
along with analyses of how they scale with the number of spins.
The quantum-mechanical nature of processes involving such
macroscopic pseudo-pure states also is a matter of debate,
and in order to discuss this issue in concrete terms
we present the results of NMR experiments which
constitute a macroscopic analogue Hardy's paradox.
Finally, a detailed product operator description is
given of recent NMR experiments which demonstrate a
three-bit quantum error correcting code, using field
gradients to implement a precisely-known decoherence model.
\end{abstract}

\section{Introduction}
It has recently proven possible to perform simple
quantum computations by liquid-state NMR spectroscopy
\cite{ChGeKuLe:98,ChVaZhLeLl:98,CorFahHav:97,CMPKLZHS:98,%
CorPriHav:98,GershChuan:97,JonMosHan:98,NieKniLaf:98}.
This unprecedented level of coherent control promises
to be quite useful not only in demonstrating the validity
of many of the basic ideas behind quantum information processing,
but more importantly, in providing researchers in the field
with new physical insights and concrete problems to study.
This is particularly true since the ensemble nature of the
systems used for NMR computing differs substantially from the
systems previously considered as candidate quantum computers.
The use of ensembles provides tremendous redundancy, which
makes computation with them relatively resistant to errors.
It also has the potential to provide access to a limited
form of massive classical parallelism \cite{CorFahHav:97},
which could for example be used to speed up searches
with Grover's algorithm by a constant but very large
factor \cite{BoBrHoTa:98,Grover:97a,LinBarFre:98}.
The barriers that have been encountered in extending NMR
computing to nontrivial problems further raise interesting
questions regarding the relations between microscopic
and macroscopic order, and between the quantum and
classical worlds \cite{GiuliniEtAl:96,Peres:95}.

NMR computing is also contributing to quantum
information processing through the assimilation
of theoretical and experimental NMR techniques.
These techniques have been developed
over half a century of intensive research,
and grown so advanced that a recent book on the subject
is entitled ``Spin Choreography'' \cite{Freeman:98}.
It is noteworthy that, due to the scope of its applications,
NMR is now more often studied in chemistry and
even biochemistry than it is in physics and
engineering, where it was initially developed.
This has had the effect that a large portion of these
techniques have been discovered empirically and put
into the form of intuitive graphical or algebraic rules,
rather than developed mathematically from well-defined principles.
Thus the interest which NMR computing is attracting from
the quantum information processing side likewise has
the potential to benefit the field of NMR spectroscopy,
particularly through the application of algorithmic,
information theoretic and algebraic techniques.

Finally, NMR has the potential to contribute in
significant ways to the development of its own mathematics,
in the same ways that computers have contributed to the
development of recursive function theory, number theory,
combinatorics and many other areas of mathematics.
By performing experiments which can be interpreted as
computations in homomorphic images on the algebras
that are naturally associated with NMR spectroscopy,
it may be possible to obtain insights into,
or even ``proofs'' of, algebraic properties
that would otherwise be inaccessible.
To give some idea of its potential computational power,
we point out that the spin dynamics of a crystal
of calcium fluoride one millimeter on a side,
which can be highly polarized, superbly
controlled and measured in microscopic detail by
NMR techniques \cite{TangWaugh:92,ZhangCory:98},
is described by an exponential map in an algebra on
about $4^{10^{11}}$ physically distinct dimensions.

This paper will survey our recent research on quantum
information processing by liquid-state NMR spectroscopy,
including some new experiments which serve
to clarify the underlying principles.
We begin with a geometric interpretation of the quantum
mechanical states and operators of an ensemble of
identical spin $1/2$ particles, both pure and mixed,
which provides considerable insight into NMR.
The corresponding geometric algebra is then
extended to the {\em product operator formalism\/},
which is widely used in analyzing NMR experiments,
and which constitutes a facile framework within which to study
quantum information processing more generally \cite{SomCorHav:98}.
We proceed to use this formalism to give an
overview of the basic ideas behind ensemble quantum
computing by liquid-state NMR spectroscopy,
with emphasis on ``pseudo-pure'' state preparation and scaling.
Next, we consider one way in which quantum correlations can
appear to be present even in weakly polarized spin ensembles,
and illustrate this with the results of NMR
experiments which constitute a macroscopic
analogue of Hardy's paradox \cite{Hardy:93}.
Finally, the utility of NMR and its associated
product operator formalism as a means of studying
decoherence will be demonstrated by an analysis
of our recent experiments with a three-bit quantum
error correcting code \cite{CMPKLZHS:98}.\footnote{
The reader is assumed throughout to be familiar
with the basic notions of quantum information
processing, as presented in e.g.\ Refs.~\cite{%
Peres:95,Preskill:98,Steane:98,WilliClear:98}.
Excellent detailed expositions of NMR spectroscopy are also available,
see e.g.\ Refs.~\cite{ErnBodWok:87,Freeman:98,Munowitz:88,Slichter:90}.
A more introductory account of our work on ensemble quantum
computing by NMR spectroscopy, directed primarily towards
physicists, may be found in Ref.\ \cite{CorPriHav:98}. }

\vspace*{0.20in}
\section{The geometry of spin states and operators}
NMR spectroscopy is based on the fact that
the nuclei in many kinds of atoms are endowed
with an intrinsic angular momentum, the properties
of which are determined by an integer or half-integer
quantum number $S \ge0$, called the nuclear {\em spin\/}.
For the purposes of quantum information processing by NMR,
it will suffice to restrict ourselves to spin $S = 1/2$.
In this case, measurement of the component
of the angular momentum along a given axis in
space always yields one of two possible values:
$\pm\hbar/2$ (where $\hbar$ is Planck's constant $h$ over $2\pi$).
According to the principles of quantum mechanics,
the quantum state of the ``spin'' (nucleus) after
such a measurement may be completely characterized
by one of two orthonormal vectors in a two-dimensional
Hilbert (complex vector) space $\cal H$, with Hermitian
(sesquilinear) inner product $\AVG{\cdot|\cdot}$.
A rotation of this axis in physical space
induces a transformation in $\cal H$ by an
element of the special unitary group $\LAB{SU}(2)$,
which is the two-fold universal covering group of the
three-dimensional Euclidean rotation group $\LAB{SO}(3)$,
and the elements of $\cal H$ are called
{\em spinors\/} to emphasize this fact.
The Lie algebra basis $(\IX,\IY,\IZ)$ of $\LAB{SU}(2)$
(or $\LAB{SO}(3)$) corresponding to infinitesmal rotations about
three orthogonal axes in space satisfies the commutation relations
\begin{equation} \label{eq:com_rel}
\left[\IX,\IY\right] ~=~ \imath \IZ ~,\quad
\left[\IZ,\IX\right] ~=~ \imath \IY ~,\quad
\left[\IY,\IZ\right] ~=~ \imath \IX ~,
\end{equation}
and the eigenvalues $\pm1/2$ of these three Hermitian
(self-adjoint) operators correspond to the possible
outcomes of measurements of the angular momentum
(in units of $\hbar$) along the three axes.\footnote{
Detailed explanations of these basic features of the
quantum mechanics of spin may be found in modern textbooks.
We would particularly recommend Sakurai \cite{Sakurai:94},
for an introduction to the underlying physics,
or the monograph by Biedenharn and Louck \cite{BiedeLouck:81},
for a complete mathematical development. }

The Hilbert space representation of the
kinematics of an isolated spin, however,
is not sufficient to describe the joint state
of the macroscopic collections of spins
which are the subject of NMR spectroscopy.
A {\em mixed state\/} (as opposed to the
{\em pure state\/} of an isolated spin) is a random
{\em ensemble\/} of spins not all in the same pure state.
This ``ensemble'' could be a thought-construction
which describes our state-of-knowledge of a single spin
(as used in J.~W.~Gibbs' formulation of statistical thermodynamics),
or it could be a very large physical collection of spins,
as in an NMR sample tube.
In either case, a probability is assigned to every possible spinor,
which can be interpreted as its frequency of occurrence in the ensemble
(but see Ref.\ \cite{Jaynes:57} for a Baysian point-of-view).
The Heisenberg uncertainty principle limits what can
be known about the ensemble to the {\em ensemble-average
expectation values\/} of the quantum mechanical observables.
This information, in turn, can be encoded into a single
operator on $\cal H$, called the {\em density operator\/}.

To define this operator mathematically,
we first recall the canonical algebra isomorphism
between the endomorphisms $\FUN{End}({\cal H})$
and the tensor product ${\cal H} \otimes {\cal H}^*$
of $\cal H$ with its dual space ${\cal H}^*$.
Denoting the dual of a vector $\KET{\psi}$ under
the Hermitian inner product of $\cal H$ by $\BRA{\psi}$,
the composition product in $\FUN{End}({\cal H})$
corresponds to a product on ${\cal H} \otimes {\cal H}^*$
which is given on the factorizable tensors by
\begin{equation}
(\KET{\varphi} \otimes \BRA{\varphi'})
(\KET{\vartheta} \otimes \BRA{\vartheta'})
~=~ \AVG{\varphi'\,|\,\vartheta} \,
(\KET{\varphi} \otimes \BRA{\vartheta'}) ~,
\end{equation}
and extended to all tensors by linearity.
Following common practice, we shall usually
drop the tensor product sign ``$\otimes$''
and write this {\em dyadic product\/}
as $\KET{\varphi}\BRA{\vartheta}$.
The restriction of this product to the diagonal,
$\KET{\psi}\BRA{\psi}$, linearly spans the (real) subspace
of all Hermitian operators in $\FUN{End}({\cal H})$,
and the action of $\LAB{SU}(2)$ on these products
is its usual action on such operators,
\begin{equation}
\KET{\psi}\BRA{\psi} \quad\mapsto\quad
\VEC U\, \KET{\psi}\BRA{\psi} \,\tilde{\VEC U} ~,
\end{equation}
where $\tilde{\VEC U} \equiv \VEC U^\sim$ denotes the
Hermitian conjugate (adjoint) of $\VEC U \in \FUN{SU}(2)$.

Restricting ourselves to an ensemble
involving a finite set of states
$\{\KET{\psi_k}\}$ for ease of presentation,
the density operator may now be defined as \cite{Blum:96}
\begin{equation}
\RHO ~\equiv~ \overline{\KET{\psi}\BRA{\psi}}
~\equiv~ {\sum}_k\, p_k\, \KET{\psi_k}\BRA{\psi_k} ~,
\end{equation}
where the $p_k \ge0$ are the probabilities of the
various states in the ensemble ($\sum_k p_k = 1$).
Because $\BRA{\varphi}\,\RHO\,\KET{\varphi} = \sum_k p_k\,
|\AVG{\varphi\,|\,\psi_k}|^2 \ge0$ for any spinor $\KET{\varphi}$,
the density operator is necessarily positive semi-definite.
Letting ``$\TR$'' be the contraction operation on
${\cal H} \otimes {\cal H}^*$ (or trace on $\FUN{End}({\cal H})$),
letting $\VEC A \in \FUN{End}({\cal H})$ be any
Hermitian operator, and using the invariance of
the trace under cyclic permutations, we find that
\begin{equation}
\TR\left( \VEC A\,\RHO \right) ~=~ 
{\sum}_k\, p_k\, \TR\left(\,\VEC A\, \KET{\psi_k}\BRA{\psi_k}\,\right)
~=~ {\sum}_k\, p_k\, \BRA{\psi_k} \, \VEC A\, \KET{\psi_k} ~.
\end{equation}
This proves our claim that all ensemble-average
expectation values can be obtained from $\RHO$.
Note in particular that $\TR(\RHO) = 1$.

Before showing how this applies to NMR spectroscopy,
we wish to introduce an important geometric
interpretation of the spin $1/2$ density operator,
and indeed of the entire operator algebra.
As operators, the angular momentum components
transform under $\LAB{SU}(2)$ by conjugation,
i.e.~$\VEC I_w \mapsto \VEC{UI}_w\tilde{\VEC U}$
($w \in \{\LAB{x},\LAB{y},\LAB{z}\}$).
Thus $\VEC U$ and $-\VEC U$ induce the same transformation,
so that conjugation constitutes a transitive group action of
$\LAB{SO}(3)$ on the real linear space $\AVG{\IX, \IY, \IZ}$.
It follows that this space is naturally regarded
as a three-dimensional Euclidean vector space.
Note further that $\AVG{\II, \IX, \IY, \IZ}$ equals the
four-dimensional space of Hermitian operators on ${\cal H}$,
where $\II$ is the identity on ${\cal H}$ which we
will henceforth identify with the scalar identity $1$.
This shows that any density operator can be uniquely
expanded as the sum of a {\em scalar\/} and a {\em vector\/}:
\begin{equation}
\RHO ~=~ \HALF\,\TR(\RHO)\, + \,\TR(\IX\,\RHO)\,2\IX\,
+ \,\TR(\IY\,\RHO)\,2\IY\, + \,\TR(\IZ\,\RHO)\,2\IZ\,
~\equiv~ \HALF( 1 + \VEC P )
\end{equation}
We call $\VEC P$ the {\em polarization vector\/},
since its length $P \equiv \|\VEC P\| \le1$
(the polarization) is a measure of the overall degree
of alignment of the spins in the ensemble along $\VEC P$.
The positive semi-definiteness of $\RHO$ implies $P \le1$,
and if $P = 1$, the density operator describes a (ensemble of
spins in the same) pure state up to an overall phase factor.
In this latter case the density operator can be written as
$\RHO = \KET{\psi}\BRA{\psi}$ for some spinor $\KET{\psi}$,
and hence is {\em idempotent\/} (equal to its square).
This vectorial interpretation of two-state quantum
systems became widely known through the work of
Feynman, Vernon and Helwarth \cite{FeyVerHel:57},
although it is inherent in the phenomenological equations for
NMR first proposed by F.~Bloch \cite{Bloch:46} (see below).

To extend this geometric interpretation to the
entire algebra generated by $\AVG{\II, \IX, \IY, \IZ}$,
we regard the composition product of angular momentum
operators as an associative bilinear product of vectors.
We shall call this the {\em geometric vector product\/}.
Since the eigenvalues of the $(S = 1/2)$ angular momentum operators
are $\pm1/2$, the eigenvalues of their squares are both $1/4$,
from which it follows that $(2\IX)^2 = (2\IY)^2 = (2\IZ)^2 = 1$.
In accord with the isotropy of space, moreover,
\begin{equation}
(\VEC U\VEC I_w\tilde{\VEC U})^2 ~=~
\VEC U (\VEC I_w)^2 \tilde{\VEC U} ~=~ 1/4
\end{equation}
for all $\VEC U \in \LAB{SU}(2)$ and
$w \in \{\LAB{x},\LAB{y},\LAB{z}\}$,
which together with the bilinearity of the
product implies that the square of {\em any\/}
vector is equal (relative to the orthonormal
basis $(2\IX,2\IY,2\IZ)$) to its length squared.
Via the law of cosines, we can now show that the
{\em symmetric part\/} of the geometric product of
any two vectors is their usual Euclidean inner product:
\begin{equation} \begin{array}{rl}
\VEC A \cdot \VEC B ~= & \HALF \left( \|\VEC A\|^2
+ \|\VEC B\|^2 - \|\VEC A - \VEC B\|^2 \right) \\
= & \HALF \left( \VEC A^2 + \VEC B^2 - (\VEC A - \VEC B)^2 \right)
~=~ \HALF \left( \VEC{AB} + \VEC{BA} \right)
\end{array} \end{equation}
The commutation relations in Eq.~(\ref{eq:com_rel}), on the
other hand, show that the {\em antisymmetric part\/} is equal
(up to a factor of $-\imath$) to the usual vector cross product:
\begin{equation}
\VEC A \times \VEC B ~=~ -\FRAC{\imath}2 \left[ \VEC A, \VEC B \right]
~=~ -\FRAC{\imath}2(\VEC A \VEC B - \VEC B \VEC A)
~\equiv~ -\imath (\VEC A \wedge \VEC B)
\end{equation}
We call the antisymmetric part
$\VEC A \wedge \VEC B = \imath(\VEC A \times \VEC B)$
the {\em outer product\/} of $\VEC A$ and $\VEC B$,
and note that it is geometrically distinct from vectors
because inversion in the origin does not change it.
Such things have been called ``axial vectors'', although
we prefer the older and more descriptive term {\em bivector\/}.
On writing the geometric product as the sum of its symmetric and
antisymmetric parts, $\VEC{AB} = \VEC A\cdot\VEC B + \VEC A\wedge\VEC B$,
we see that perpendicular pairs of vectors anticommute.
It follows that the three basis bivectors $\imath2\IX$,
$\imath2\IY$ and $\imath2\IZ$ also anticommute.
These square to $-1$ rather than $1$, however,
and thus can be identified with the usual
{\em quaternion\/} units \cite{Altmann:86}.
Finally, the {\em unit pseudo-scalar\/}
$8\IX\IY\IZ$ likewise squares to $-1$,
which together with the fact that it
commutes with the basis vectors and hence
everything in the algebra enables it to
be identified with the unit imaginary
$\imath$ itself \cite{GulLasDor:93}.

This algebra is often called the {\em Clifford algebra\/}
of a three-dimensional Euclidean vector space,
although we shall use the term {\em geometric algebra\/}
here (which W.~K.~Clifford himself used).
Such an algebra is canonically associated with any
metric vector space, and provides a natural algebraic
encoding of the geometric properties of that space.
The fact that the three-dimensional Euclidean version
can be defined starting from the well-known properties
of the spin $1/2$ angular momentum operators indicates that
a large part of quantum mechanics is really just an unfamiliar
(but extremely elegant and facile \cite{Havel$ECC:98,Hestenes:86})
means of doing Euclidean geometry.
Geometric algebra has more recently been extensively advocated
and used to demystify quantum physics by a number of groups
\cite{Baylis:96,DorLasGul:93,Hestenes:66,Lounesto:97}.
Of particular interest are recent proposals to use the
geometric algebra of a direct sum of copies of Minkowski
space-time to obtain a relativistic multiparticle theory,
from which all the nonrelativistic theory used in this paper
falls out naturally as a quotient subalgebra \cite{DoLaGuSoCh:96}.

We are now ready to describe the
simplest possible NMR experiment.
The time-independent Schrodinger equation is
\begin{equation}
\imath\hbar\, \KET{\dot\psi} ~=~ \VEC H\, \KET{\psi} ~,
\end{equation}
where the Hamiltonian $\VEC H$ is the generator of
motion and the ``dot'' denotes the time derivative.
This implies that the density operator evolves according
to the {\em Liouville-von Neumann equation\/}:
\begin{equation} \begin{array}{rl}
\imath\hbar \, \dot{\RHO}
~= & \imath\hbar \, {\sum}_k\, p_k \left(
\KET{\dot\psi_k} \BRA{\psi_k} +
\KET{\psi_k} \BRA{\dot\psi_k} \right) \\
=~ & {\sum}_k\, p_k \left( \VEC H \KET{\psi_k} \rule[0pt]{0pt}{12pt}
\BRA{\psi_k} - \KET{\psi_k} \BRA{\psi_k} \VEC H \right)
~=~ \left[ \,\VEC H, \RHO\, \right]
\end{array} \end{equation}
The dominant Hamiltonian in NMR is the Zeeman
interaction of the magnetic dipoles of the spins
(which is parallel to their angular momentum vectors)
with a constant applied magnetic field $\VEC B_0$.
This {\em Zeeman Hamiltonian\/} is given by
$\VEC H_\LAB{Z} = -\HALF\gamma\hbar\VEC B_0$,
where $\gamma$ is a proportionality constant
called the {\em gyromagnetic ratio\/}, which together with
the above gives the {\em Bloch equation\/} \cite{Bloch:46}:
\begin{equation}
\dot{\VEC P} ~=~ \dot{\RHO}
~=~ -\imath \HALF \gamma \left[\, \RHO, \VEC B_0\, \right]
~=~ \gamma\, \VEC P \times \VEC B_0
\end{equation}
The solution to this equation is $\RHO(t) = \VEC U \RHO(0)\,
\tilde{\VEC U}$ with $\VEC U = \exp(-\imath t \VEC H_\LAB{Z})$,
which is a time-dependent rotation of the polarization
vector about the magnetic field with a constant angular
velocity $\omega_0 \equiv \gamma\hbar\,\|\VEC B_0\|$.
This ``classical'' picture is
an example of Ehrenfest's theorem,
and is analogous to the precession of
a gyroscope in a gravitational field.

Throughout this paper we adopt the universal
convention that the magnetic field is along
the $\LAB{z}$-axis: $\VEC B_0 = B_0 2\IZ$.
The component of the net precessing magnetic moment of
the spins in the transverse $\LAB{xy}$-plane generates
a complex-valued radio-frequency electrical signal
proportional to $\TR((\IX + \imath\IY)\RHO(t)) =
2\IX \cdot \VEC P(t) + \imath 2\IY \cdot \VEC P(t)$,
whose Fourier transform is an NMR spectrum
containing a peak at the precession frequency of
each distinct kind of spin present in the sample.
This has the important consequence that in NMR we measure
the {\em expectation values\/} of the observables directly,
which is due in turn to the fact that we are measuring
the sum of the responses of the spins over the ensemble.
These measurements yield negligible information on the
quantum state of the individual spins in the ensemble
and hence are nonperturbing, in that they do not
appreciably change the state of the ensemble as a whole.
Such {\em weak measurements\/} contrast starkly with the strong
measurements usually considered in quantum mechanics,
where determining the component of a spin along
an axis yields one of two possible values and
``collapses'' it into the corresponding basis state,
so that only one classical bit of information
can be obtained \cite{Peres:95,Sakurai:94}.
A discussion of the computational implications of weak
measurements may be found in Ref.\ \cite{CorFahHav:97}.

The natural (minimum energy) orientation of the spins'
dipoles in a magnetic field is parallel to the field,
and thus to obtain a precessing magnetic dipole
it is necessary to rotate the polarization
vector $\VEC P$ away from the field axis $2\IZ$.
This is done by applying an additional, rotating
magnetic field $\VEC B_1$ of magnitude $B_1$ in the
$\LAB{xy}$-plane perpendicular to the static field $\VEC B_0$,
which gives the time-dependent Hamiltonian
\begin{equation}
\VEC H ~=~ \VEC H_\LAB{Z} + \VEC H_\LAB{RF}
~=~ -\gamma\hbar \left( B_0 \IZ + B_1
(\cos(\omega t)\IX + \sin(\omega t)\IY) \right) ~.
\end{equation}
The effect of such a rotating field is most
easily determined by transforming everything
into a frame which rotates along with it,
in which the Hamiltonian becomes time-independent:
\begin{equation}
\RHO' ~=~ e^{-\imath\omega t\IZ} \RHO\, e^{\imath\omega t\IZ} ~,\quad
\VEC H' ~=~ e^{-\imath\omega t\IZ} \VEC H e^{\imath\omega t\IZ}
~=~ \VEC H_\LAB{Z} + \gamma\hbar B_1 \IX
\end{equation}
Then the Bloch equation itself is transformed as follows:
\begin{equation} \begin{array}{rl}
{\rule[0pt]{0pt}{7pt}\smash{\dot{\VEC P}}}' ~= & -\imath\omega \IZ \VEC P'
+ e^{-\imath\omega t\IZ} \dot{\VEC P} \,
e^{\imath\omega t\IZ} + \VEC P' \imath\omega \IZ \\
= & \VEC P' \times (\VEC H' - \omega \IZ)
\label{eq:trans_bloch}
\end{array} \end{equation}
Thus if $\omega$ equals the natural precession
frequency of the spins $\omega_0 = \gamma\hbar B_0$,
the Zeeman Hamiltonian $\VEC H_\LAB{Z}' =
\VEC H_\LAB{Z} = \omega_0 \IZ$ cancels out.
In this frame, the spins turn about a (rotating) axis
perpendicular to $\VEC B_1'$ at a rate $\omega_1 = \gamma\hbar B_1$,
so that if the polarization vector starts out along $\LAB{z}$,
it is in the $\LAB{xy}$-plane where it produces
the maximum signal after a time $t=\pi/(2\omega_1)$.
Henceforth, all our coordinate frames will be rotating
at the transmitter frequency unless otherwise mentioned.

\vspace*{0.20in}
\section{The product operator formalism}
Thus far we have restricted our presentation to
ensembles consisting of indistinguishable nuclear spins.
The power of NMR spectroscopy as a means of chemical analysis,
however, depends on the fact that the different nuclei in
a molecule generally have distinct electronic environments,
which affect the applied magnetic field at each nucleus.
As a result, they precess at slightly different frequencies and
give rise to resolvable ``peaks'' in the resulting spectrum.
This is also one of the reasons why NMR provides a
facile approach to quantum information processing,
since it permits each {\em chemical\/} equivalence class
of spins in the ensemble to be treated as a separate ``qubit''.
In this section we will describe an extension
of the density operator to multispin systems,
using a basis which is a direct generalization of the
``scalar + vector'' basis given above for a single spin.
We then illustrate this so-called {\em product operator formalism\/}
\cite{BoulaRance:94a,ErnBodWok:87,SomCorHav:98,SoEiLeBoEr:83,vdVenHilbe:83}
by describing how quantum information processing can be done on an ensemble
of multispin molecules, using the internal Hamiltonian of liquid-state NMR.
For the sake of simplicity we shall assume throughout
that the ensemble is in a pure state, i.e.\ that the
joint state of the spins in every molecule is identical.
The next section is devoted to the complications
involved in extending this approach to the highly
mixed states which are available in practice.

As usual in quantum information
processing \cite{Steane:98,WilliClear:98},
we choose a {\em computational basis\/}
$(\KET{0},\KET{1})$ for the Hilbert space $\cal H$ of each spin
that corresponds to the eigenvectors of its $\IZ$ operator,
i.e.\ to alignment of the spin with (up) and against (down)
a magnetic field $\VEC B_0$ along the $\LAB{z}$ axis.
Relative to this basis, a superposition $c_0\KET{0} + c_1\KET{1}$
($c_0, c_1 \ne 0$ complex with $|c_0|^2 + |c_1|^2 = 1$)
is any state with transverse ($\LAB{xy}$) components.
The Hilbert space needed to describe the kinematics
of a system consisting of $N$ distinguishable spins
({\em not\/} an ensemble) is the $N$-fold tensor product
of their constituent Hilbert spaces \cite{Peres:95,Sakurai:94}.
The induced basis in this $(2^N)$-dimensional space is
\begin{equation}
\KET{\kappa^1} \otimes \KET{\kappa^2} \otimes\cdots\otimes \KET{\kappa^N}
~\equiv~ \KET{\kappa^1\kappa^2\ldots\kappa^N} ~\equiv~ \KET{k} ~,
\end{equation}
where $\kappa^n \in \{ 0, 1 \}$ $(n = 1,\ldots,N)$ is the
binary expansion of the integer $k \in \{ 0,\ldots,2^N-1 \}$.
Because of the canonical isomorphism
\begin{equation}
\FUN{End}({\cal H}) \otimes \FUN{End}({\cal H})
~\approx~ \FUN{End}({\cal H} \otimes {\cal H})
\end{equation}
together with our previous isomorphism
$\FUN{End}({\cal H}) \approx {\cal H} \otimes {\cal H}^*$,
this implies that the density operators for an ensemble of $N$-spin
molecules are all contained in the $N$-fold tensor product space
\begin{equation}
({\cal H} \otimes\cdots\otimes {\cal H}) \otimes
({\cal H}^* \otimes\cdots\otimes {\cal H}^*) ~\approx~
({\cal H} \otimes {\cal H}^*) \otimes\cdots\otimes
({\cal H} \otimes {\cal H}^*) ~.
\end{equation}
It follows that a basis for the algebra of $N$-spin operators is
\begin{equation} \label{eq:bad_basis}
\begin{array}{rl}
\KET{k} \BRA{\ell} ~= &
\KET{\kappa^1\,\kappa^2\,\ldots\,\kappa^N}
\BRA{\lambda^1\,\lambda^2\,\ldots\,\lambda^N} \\
= & (\KET{\kappa^1}\BRA{\lambda^1}) \otimes
(\KET{\kappa^2}\BRA{\lambda^2}) \otimes\cdots\otimes
(\KET{\kappa^N}\BRA{\lambda^N}) ~,
\end{array} \end{equation}
where $\kappa^n, \lambda^n \in \{ 0, 1 \}$
($n = 1,\ldots,N$) are binary expansions of
the integers $k, \ell \in \{ 0,\ldots,2^N-1 \}$.
This basis, however, does not consist of Hermitian operators,
and although the dyadic products $\KET{\psi}\BRA{\psi}$
($\KET{\psi} \in {\cal H} \otimes\cdots\otimes {\cal H}$)
do span the real subspace of all Hermitian operators,
the restriction of the basis in Eq.~(\ref{eq:bad_basis})
to the diagonal does not.

An algebra basis which has the advantage of also
being a linear basis for the subspace of Hermitian
operators is known as the {\em product operator basis\/}.
It is induced by the one-spin basis $(\II,\IX,\IY,\IZ)$,
and consists simply of the tensor products of the
angular momentum operators of the individual spins.
In the case of two spins, this basis has sixteen elements:
\renewcommand{\arraystretch}{1.1}%
\begin{equation} \begin{array}{cccc}
\II \otimes \II & \II \otimes \IX & \II \otimes \IY & \II \otimes \IZ \\
\IX \otimes \II & \IX \otimes \IX & \IX \otimes \IY & \IX \otimes \IZ \\
\IY \otimes \II & \IY \otimes \IX & \IY \otimes \IY & \IY \otimes \IZ \\
\IZ \otimes \II & \IZ \otimes \IX & \IZ \otimes \IY & \IZ \otimes \IZ
\end{array} \end{equation}
\renewcommand{\arraystretch}{1.3}%
As before, a notation which eliminates the need
for repetitive ``$\otimes$'' symbols is preferred.
This is obtained by using superscripts
for the spin indices in the operators
\begin{equation}
\VEC I_w^n ~\equiv~ \II \otimes\cdots\otimes \II \otimes
\VEC I_w \otimes \II \otimes\cdots\otimes \II \qquad
\end{equation}
($\VEC I_w$ in the $n$-th place, $n = 1,\ldots,N$,
$w \in \{\LAB{x},\LAB{y},\LAB{z}\}$),
and noting that by the {\em mixed product formula\/}
between the operator composition and tensor products:
\begin{equation}
\VEC I_u^m \VEC I_v^n
~=~ \II \otimes\cdots\otimes \VEC I_u
\otimes \II \otimes\cdots\otimes \II
\otimes \VEC I_v \otimes\cdots\otimes \II
~=~ \VEC I_v^n \VEC I_u^m
\end{equation}
($\VEC I_u$ in the $m$-th place, $\VEC I_v$ in the $n$-th,
$\,m, n = 1,\ldots,N$ with $m < n$,
and $u,v \in \{\LAB{x},\LAB{y},\LAB{z}\}$).
In the following, we will again identify the identity operator
$\II \otimes\cdots\otimes \II$ with the scalar identity $1$.
We will also be using the operator norm $\|\VEC I_w^n\|^2
\equiv \AVG{(\VEC I_w^n)^2} = (\VEC I_w^n)^2 = 1/4$ obtained
from the scalar part, rather than the more usual Frobenius norm
$\|\VEC I_w^n\|_\LAB{F}^2 = \TR((\VEC I_w^n)^2) = 2^{N-2}$
on $\FUN{End}(\cal H)$, because the former is independent of $N$.
The normalization of our basis to $\|\VEC I_w^n\| = 1/2$
rather than $1$ will be seen to have both advantages and
disadvantages, but the convention is well-established in NMR.

Just as with an ensemble consisting of a single type
of spin, a pure state may be characterized by the
idempotence of its density operator: $\RHO^2 = \RHO$.
The scalar part of the density operator is
$\AVG{\RHO} = 2^{-N} \TR(\RHO) = 2^{-N}$,
and if we write an arbitrary density operator
$\RHO \equiv \OL{\KET{\psi}\BRA{\psi}}$ in diagonal form as
\begin{equation}
\RHO ~=~ \VEC U \left( {\sum}_{k=0}^{2^N-1}\,
p_k\, \KET{k}\BRA{k} \right) \tilde{\VEC U}
\end{equation}
($0 \le p_k \le 1$, $\sum_k p_k = 1$)
for some $\VEC U \in \LAB{SU}(2^N)$,
we see that the idempotence of $\RHO$ is equivalent
to $\AVG{\RHO^2} = 2^{-N}$, i.e.\ $p_\ell = 1$
for some $\ell \in \{ 0,\ldots,2^N-1 \}$.
This shows that the density operator of a
pure state is in fact a primitive idempotent.
Without loss of generality we may take $\ell = 0$, so that
$\RHO = \VEC U \KET{00\ldots0}\BRA{00\ldots0} \tilde{\VEC U}$.
If we expand $\KET{0}\BRA{0}$ in the product operator
basis, we obtain $\EP \equiv \HALF(1 + 2\IZ)$,
and similarly for $\KET{1}\BRA{1} = \EM \equiv \HALF(1 - 2\IZ)$.
Thus we can also write the density operator of a pure state as
\begin{equation}
\RHO ~=~ \VEC U \left( \EP^1 \EP^2 \cdots \EP^N \right) \tilde{\VEC U} ~,
\end{equation}
where the superscript on the idempotent
$\EP$ is the spin index as before.
More generally, the set of all density operators
consists of the closed convex cone of positive
semi-definite operators in the Hermitian subspace of
$\FUN{End}( {\cal H} \otimes\cdots\otimes {\cal H} )$,
and the density operators of pure states
are the extreme rays of this cone.

We now consider the form of the Hamiltonian
which is operative among the spins of an
ensemble of molecules in the liquid state
(with which we are exclusively concerned in this paper),
again using the product operator formalism.
First, there is the Zeeman Hamiltonian
previously given for a single spin, i.e.
\begin{equation}
\VEC{H}_\LAB{Z} ~\equiv~ -\omega_0^1 \IZ^1 -\cdots- \omega_0^N \IZ^N
\end{equation}
with $\omega_0^n = \hbar \gamma^n (1 - \sigma^n) B_0$,
where $\gamma^n$ the gyromagnetic ratio of the $n$-th spin
and $0\le\sigma^n\le1$ is the {\em chemical shift\/}
due to the (usually small) influence of the electronic
environment of the spins on their precession frequencies.
This Hamiltonian is easily seen to be diagonal in the
computational basis $\KET{k}$ ($k = 0,\ldots,2^N-1$),
with eigenvalues $(\pm\omega_0^1 \pm\cdots\pm \omega_0^N)/2$.

Second, there is an exchange interaction
known as the $J$ or {\em scalar coupling\/},
which is proportional to the inner product
of the spins' polarization vectors, namely
\begin{equation}
\VEC H_\LAB{J} ~=~ {\sum}_{m,n}\, 2\pi J^{mn}
\left( \IX^m\IX^n + \IY^m\IY^n + \IZ^m\IZ^n \right) ~,
\end{equation}
where $J^{mn}$ is the coupling strength in Hertz.
This interaction is mediated by the electrons in
the chemical bonds between atoms, and is usually
negligible for atoms separated by more than three bonds.
Standard perturbation theory \cite{Sakurai:94}
shows that the eigenvalues of the total Hamiltonian
$\VEC H = \VEC H_\LAB{Z} + \VEC H_\LAB{J}$ are
given to first order by the diagonal elements of
\begin{equation}
\VEC H' ~=~ \VEC H_\LAB{Z} + \VEC H_\LAB{J}' ~\equiv~
\VEC H_\LAB{Z} + 2\pi\, {\sum}_{m,n}\, J^{mn} \IZ^m\IZ^n ~,
\end{equation}
whereas the eigenvectors are given to first order by:
\begin{equation}
\KET{\ell}' ~=~ \KET{\ell} + 2\pi\, {\sum}_{m,n}\, J^{mn}\, {\sum}_{k\ne\ell}\,
\frac{\BRA{k}\, \IX^m\IX^n + \IY^m\IY^n + \IZ^m\IZ^n \,\KET{\ell}}
{\BRA{\ell}\,\VEC H_\LAB{Z}\,\KET{\ell} - \BRA{k}\,\VEC H_\LAB{Z}\,\KET{k}}
\, \KET{k}
\end{equation}
The numerator of each term in the summations
is nonzero only if $\kappa^p = \lambda^p$ for
$p \ne m, n$ and $\kappa^m = (1 - \lambda^n)$,
in which case it is $\pi J^{mn}$, while
the denominators of the corresponding
terms are $\omega_0^m - \omega_0^n$.
It follows that the eigenvectors are negligibly
perturbed so long as the frequency {\em differences\/}
are much larger than the scalar couplings,
i.e.\ $|\omega_0^m - \omega_0^n| \gg \pi|J^{mn}|$.
We shall be making this {\em weak coupling\/}
approximation throughout.

Another, potentially quite large term in the molecular spin Hamiltonian
is a through-space interaction between the spins' magnetic dipoles.
Because of the rapid motions of the molecules in a liquid,
however, these interactions are averaged to zero much
more quickly than they can have any net effect.
The effective absence of this interaction
nevertheless has the important consequence that the
{\em spins in different molecules do not interact\/},
and hence cannot be correlated with one another.\footnote{
More precisely, the spins do not interact to
an excellent, but first-order, approximation;
second-order effects do exist and are a source
of spin-spin relaxation (aka decoherence).}
As a result, the density operator of the entire sample
$\VRHO$ (which describes an abstract Gibbs ensemble
obtained by tracing over the spins' environment)
can be factorized into a product of density
operators for the individual molecules, i.e.
\begin{equation}
\VRHO ~=~ \VRHO^1 \cdots \VRHO^M ~=~
\RHO^1 \otimes\cdots\otimes \RHO^M ~,
\end{equation}
where $\VRHO^m \equiv \II \otimes \cdots
\otimes \RHO^m \otimes \cdots \otimes \II$.
In a pure liquid (or if we are looking
at just one component of a solution),
all the molecules are equivalent so that
all these density operators are the same.
It follows that we can work with the partial
trace over all but any one of the molecules,
which is called the {\em reduced\/} density
operator $\RHO ~\equiv~ \RHO^1 ~=\cdots=~ \RHO^M$.
Since this operates on a space of dimension $2^N$ where
$N$ is now the number of spins in a single molecule,
rather than $2^{MN}$ where $M \sim 10^{20}$
is the number of molecules in the sample,
this is a very considerable simplification.
It also means that in liquid-state NMR we are
working with a physical ensemble (the sample),
rather than a purely abstract Gibbs ensemble.

Finally, there is the interaction of the spins
with a transverse RF (radio-frequency) field,
which we described in the last section.
Whenever weak coupling is valid,
we can apply this field in a single ``pulse'',
tuned to the precession frequency of the $k$-th spin (say),
which is short enough that we may neglect the evolution
of the spins due to scalar coupling while it lasts.
This effects a unitary transformation of the form
\begin{equation} \begin{array}{rcl}
e^{-\imath\theta\IX^n} &=& 1 -
\imath \left(\FRAC{\theta}{2}\right) 2\IX^n -
\HALF \left(\FRAC{\theta}{2}\right)^2 +
\FRAC{\imath}{6} \left(\FRAC{\theta}{2}\right)^3
\,2 \IX^n + \cdots \\ &=&
\cos\left(\FRAC{\theta}{2}\right) -
\imath \sin\left(\FRAC{\theta}{2}\right) 2\IX^1 ~,
\end{array} \end{equation}
which corresponds to a right-hand rotation
of the $k$-th spin by an angle $\theta$ about
the $\LAB{x}$ axis in the rotating frame.
Using a pulse with a broad frequency range,
it is also possible (in fact easier) to apply
such a rotation to all the spins in parallel.

We will now indicate how RF pulses, in combination
with the innate Hamiltonian of the spins, enable us
to implement standard quantum logic gates in a manner
similar to that considered by computer scientists studying
universality in quantum computation \cite{BBCDMSSSW:95}.
The simplest such gate is the NOT operation on the
e.g.\ first spin, which simply rotates it by $\pi$;
combining the above formula with the basic geometric
algebra relations $\IX\IZ = -\IZ\IX$ and $(2\IX)^2 = 1$,
we obtain
\begin{equation} \begin{array}{rcl}
e^{-\imath\pi\IX^1} \EP^1 e^{\imath\pi\IX^1}
&=& (-2\imath\IX^1) \EP^1 (2\imath\IX^1)
~=~ \HALF (1 + 8 \IX^1 \IZ^1 \IX^1) \\
&=& \HALF (1 - 8 \IX^1 \IX^1 \IZ^1)
~=~ \HALF (1 - 2\IZ^1) ~=~ \EM^1 
\end{array} \end{equation}
The c-NOT (controlled-NOT) gate, on the other hand,
is a $\pi$ rotation of e.g.\ the first spin
{\em conditional\/} on the polarization of a second.
Using the relation $\EPM^2 \EMP^2 = 0$, we can easily show that
\begin{equation}
(-2\imath\IX^1 \EM^2 + \EP^2) (\VEC{E}_\epsilon^1 \EPM^2)
(2\imath\IX^1 \EM^2 + \EP^2) ~=~ \VEC{E}_{\pm\epsilon}^1 \EPM^2
\end{equation}
($\epsilon \in \{\pm{}\}$).
The phase factor $\imath$ multiplying $\IX^1$
complicates the action of the c-NOT on a superposition,
but can be eliminated by a phase shift conditional on the second spin.
Using $\EM^2 + \EP^2 = 1$, this phase-corrected
c-NOT gate is given by $\VEC{S}^{1|2} \equiv$
\begin{equation} \begin{array}{rcl}
2\IX^1 \EM^2 + \EP^2 &=&
(-\imath \EM^2 + \EP^2) (2\imath\IX^1 \EM^2 + \EP^2)
\\ &=&
\left( 1 + (e^{-\imath\frac\pi2} - 1) \EM^2 \rule{0pt}{10pt} \right)
\left( 1 + (e^{\imath\pi\IX^1} - 1) \EM^2 \right)
\\ &=&
e^{-\imath\frac{\pi}{2}\EM^2} e^{\imath\pi\IX^1\EM^2}
~=~ e^{-\imath\frac{\pi}{2}(1 - 2\IX^1)\EM^2} ~,
\end{array} \end{equation}
and hence the idempotents $\EPM^n$ also give
us an algebraic description of the c-NOT gate,
in addition to the density operators of pure states.
It is well-known that single spin rotations,
together with the c-NOT, are sufficient to
implement any quantum logic gate \cite{BBCDMSSSW:95}.

The c-NOT can be implemented in NMR by combining single
spin rotations with the conditional rotations induced
by (weak) scalar coupling $2\pi J^{12}\IZ^1\IZ^2$
\cite{CorPriHav:98,GershChuan:97,JonHanMos:98}.
Recalling that in the discrete $\LAB{SO}(3)$
subgroup of rotations by $\pi/2$,
\begin{equation}
e^{-\imath \frac{\pi}{2} \IY^1}
e^{\imath \frac{\pi}{2} \IZ^1} ~=~
e^{\imath \frac{\pi}{2} \IZ^1}
e^{\imath \frac{\pi}{2} \IX^1} ~,
\end{equation}
we can expand the above propagator as follows:
\begin{equation} \begin{array}{rcl}
\VEC S^{1|2} &=&
e^{-\imath\frac{\pi}{2}(1 - 2\IX^1)\EM^2}
\\ &=&
e^{-\imath\frac{\pi}{2}\IY^1} \, e^{-\imath\pi\EM^1\EM^2}
\, e^{\imath\frac{\pi}{2}\IY^1}
\\ &=&
\sqrt{-\imath}\, e^{-\imath\frac{\pi}{2}\IY^1}
\, e^{\imath\frac{\pi}{2}(\IZ^1+\IZ^2)}
\, e^{-\imath\pi\IZ^1\IZ^2}
\, e^{\imath\frac{\pi}{2}\IY^1}
\\ &=&
\sqrt{-\imath}\,
e^{\imath\frac{\pi}{2}(\IZ^1+\IZ^2)}
\, e^{\imath\frac{\pi}{2}\IX^1}
\, e^{-\imath\pi\IZ^1\IZ^2}
\, e^{\imath\frac{\pi}{2}\IY^1}
\end{array} \end{equation}
Since the overall phase of the transformation
has no effect on the density operator, it follows that
the c-NOT gate $\VEC{S}^{1|2}$ can be implemented by an NMR
{\em pulse sequence\/}, wherein each pulse and delay corresponds
to the indicated ``effective'' Hamiltonian in temporal order:
\begin{equation} \begin{array}{rl}
& [-\FRAC{\pi}{2}\IY^1] \rightarrow [\pi\IZ^1\IZ^2] \rightarrow
[-\FRAC{\pi}{2}\IX^1] \rightarrow [-\FRAC{\pi}{2}(\IZ^1+\IZ^2)]
\\ \Leftrightarrow
& \exp\left( \FRAC{\pi}{2}(\IZ^1+\IZ^2) \right)
\exp\left( \FRAC{\pi}{2}\IX^1 \right)
\exp\left( -\pi\IZ^1\IZ^2 \right)
\exp\left( \FRAC{\pi}{2}\IY^1 \right)
\end{array} \end{equation}
In practice, the effective Hamiltonian $[\pi\IZ^1\IZ^2]$
is obtained by applying a $\pi$-pulse to both spins in the
middle and at the end of a $1/(2J^{12})$ evolution period,
to ``refocus'' their Zeeman evolution \cite{CorPriHav:98}.
The $[-\FRAC{\pi}{2}\IY^1]$ and $[-\FRAC{\pi}{2}\IX^1]$
Hamiltonians are implemented by RF pulses as above,
while the $[-\FRAC{\pi}{2}(\IZ^1+\IZ^2)]$ transformation
is most easily implemented by letting one spin evolve
while applying a $\pi$-pulse to the other, then vice versa,
and finally realigning the transmitter's phase with the spins'.

\pagebreak
The ``readout'' procedure needed to determine the result
of an NMR computation differs somewhat that usually
considered in quantum computing \cite{Steane:98,WilliClear:98}.
The most important difference is of course the fact that
in conventional NMR one can only make weak (nonperturbing)
measurements of the observables, as previously described.
As likwise described above, these observables are the
$\LAB{x}$ and $\LAB{y}$ components $\IX^n$ and $\IY^n$
of the dipolar magnetization due to each spin in a
rotating frame defined by the transmitter frequency.
The products of the angular momentum components
of different spins (e.g.\ $\IX^1\IY^2$),
however, do not produce a net magnetic
dipole and hence cannot be detected directly.
Thus we are limited to one-spin observables,
as is usually assumed in quantum computation.
The unobservable degrees of freedom may also be
characterized in the basis $\KET{k}\BRA{\ell}$
as having a {\em coherence order\/}
$|\BRA{k}\,\IZ\,\KET{k} - \BRA{\ell}\,\IZ\,\KET{\ell}| \ne 1$,
where $\IZ \equiv \IZ^1 +\cdots+ \IZ^N$ is
the total angular momentum along $\LAB{z}$
\cite{CorPriHav:98,ErnBodWok:87,Freeman:98,Munowitz:88,Slichter:90}.

\begin{figure}
\begin{picture}(325,160)
\put(30,135){\sf Inphase}
\put(25,125){\sf Absorptive}
\put(112,135){\sf Inphase}
\put(107,125){\sf Dispersive}
\put(0,30){ \psfig{file=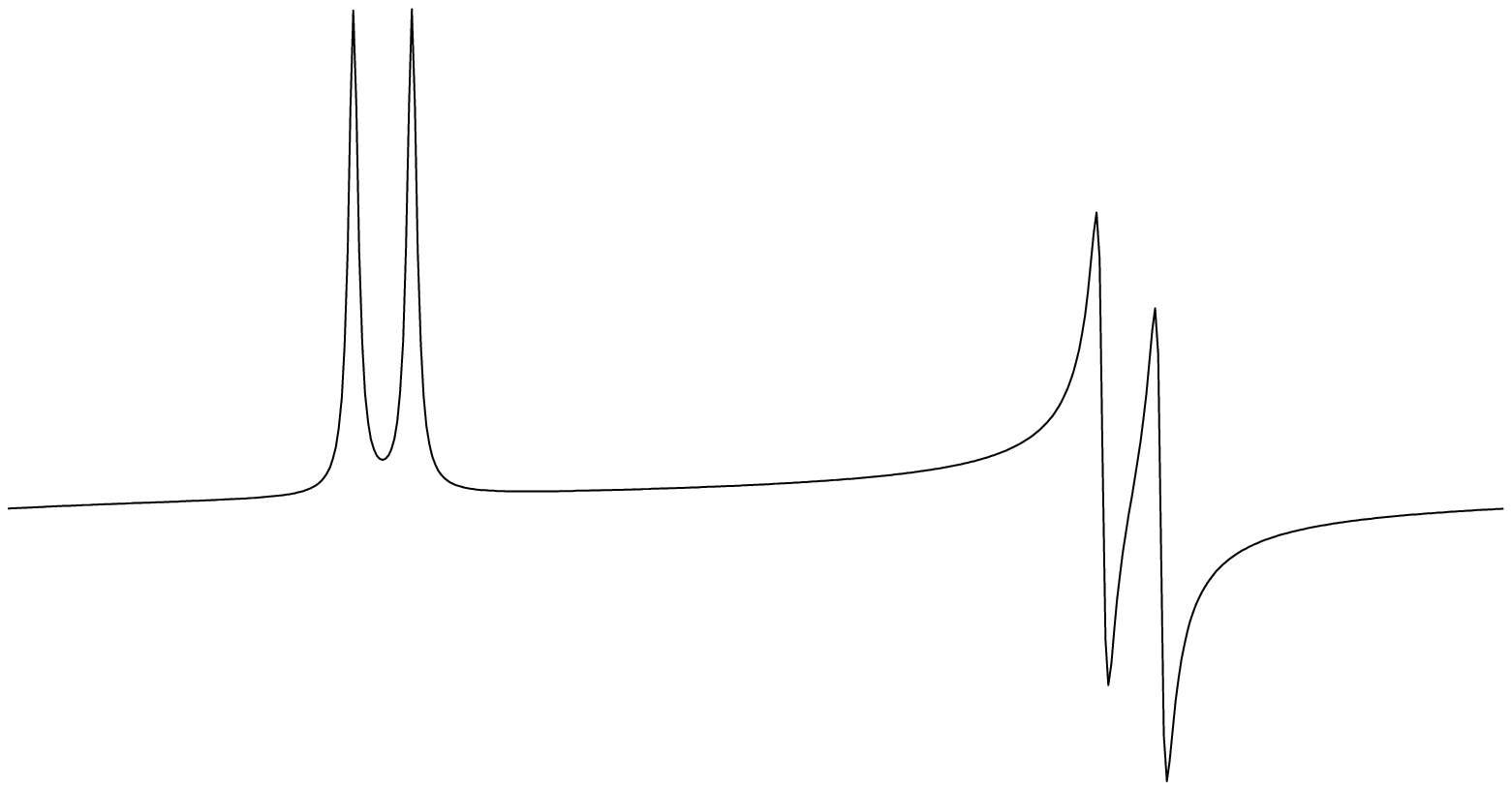,width=2.25in} }
\put(189,135){\sf Antiphase}
\put(188,125){\sf Absorptive}
\put(290,135){\sf Antiphase}
\put(290,125){\sf Dispersive}
\put(175,10){ \psfig{file=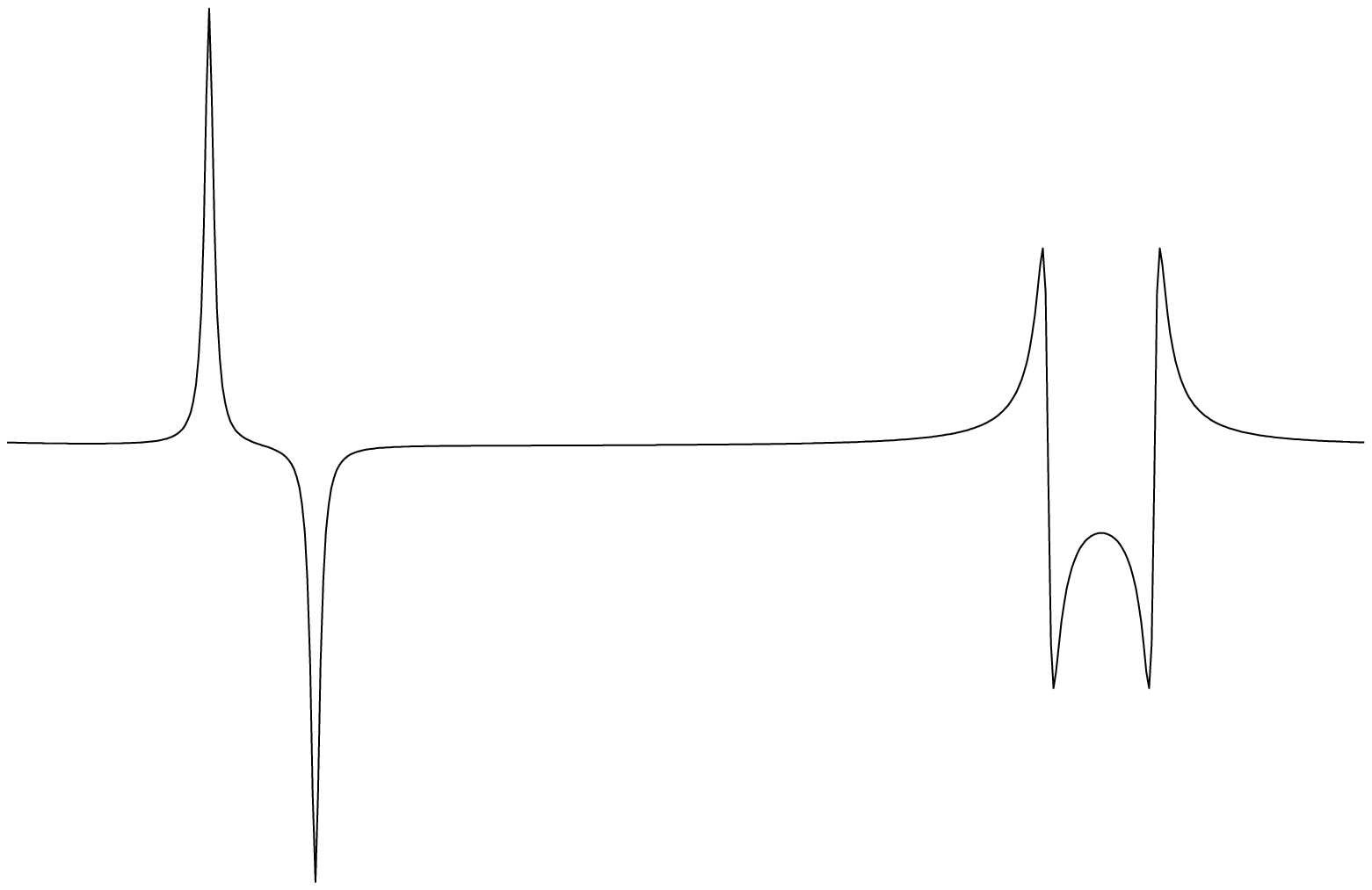,width=2.25in} }
\end{picture} \caption{
Plots of NMR spectra for a weakly coupled
two-spin molecule (amplitude versus frequency).
The left-hand plot is for the spin state
% EXPANDING THE FOLLOWING MACROS PRODUCES LINE TOO LONG IN AUX FILE
% $\IX^1 + \IY^2$,
$\mbox{\boldmath$I$}_{\sf x}^1 + \mbox{\boldmath$I$}_{\sf y}^2$,
which gives a pair of inphase absorptive peaks for spin $1$ (left)
and a pair of inphase dispersive peaks for spin $2$ (right).
The right-hand plot is for the spin state
% EXPANDING THE FOLLOWING MACROS PRODUCES LINE TOO LONG IN AUX FILE
% $\IX^1\IZ^2 + \IZ^1\IY^2$,
$\mbox{\boldmath$I$}_{\sf x}^1 \mbox{\boldmath$I$}_{\sf z}^2
+\mbox{\boldmath$I$}_{\sf z}^1 \mbox{\boldmath$I$}_{\sf y}^2$,
which gives a pair of antiphase absorptive peaks for spin $1$ (left)
and a pair of antiphase dispersive peaks for spin $2$ (right).
By fitting the peak shapes in such spectra after
various $\pi/2$ rotations of the individual spins,
one obtains sufficient information to uniquely
reconstruct the complete density operator.
} \label{fig:readout}
\end{figure}

According to the usual phase conventions of NMR,
the Fourier transform of the $\LAB{x}$-magnetization
of e.g.\ the first spin, $\IX^1$, yields an {\em absorptive\/}
peak shape, while $\IY^1$ produces a {\em dispersive\/} shape,
both centered on its precession frequency $\omega_0^1$.
If the first spin is coupled to e.g.\ the second,
its signal is modulated by $\cos(\pi J^{12} t)$
yielding a spectrum containing two peaks separated by
the coupling constant $J^{12}$ \cite{CorPriHav:98}.
An effective exception to the unobservability of
the products are those of the form $\IX^1\IZ^2$
(or $\IY^1\IZ^2$), which (when $J^{12} \ne 0$)
evolve under scalar coupling into one-spin terms.
Using the facts that $4\IZ^1\IZ^2$ and $4\IX^1\IZ^2$
anticommute while ${(4\IZ^1\IZ^2)}^2 = 1$,
we can show this as follows:
\begin{equation} \begin{array}{rcl}
e^{-\imath t \pi J^{12} 4\IZ^1\IZ^2}
\, \left( 4\IX^1\IZ^2 \right)
&=&  e^{-\imath t 2\pi J^{12} \IZ^1\IZ^2}
\, \left( 4\IX^1\IZ^2 \right) \,
e^{\imath t 2\pi J^{12} \IZ^1\IZ^2} \\
&=& \left(\cos(\pi J^{12} t) - \imath 4\IZ^1\IZ^2
\sin(\pi J^{12} t)\right) \, \left( 4\IX^1\IZ^2 \right) \\
&=& \cos(\pi J^{12} t) \, 4\IX^1\IZ^2 + \sin(\pi J^{12} t) \, 2\IY^1
\end{array} \end{equation}
Because the signal is now sinusoidally modulated by the coupling,
for a single pair of coupled spins this results in
a pair of {\em antiphase\/} peaks with opposite signs,
as opposed to the {\em inphase\/} peaks
described for $\IX^1$ and $\IY^1$ above.
These antiphase peaks may likewise be absorptive
($\IX^1\IZ^2$) or dispersive ($\IY^1\IZ^2$), respectively.
Figure \ref{fig:readout} shows examples of all these
possibilities for a pair of two coupled spins.

More generally, if the $n$-th spin is coupled to $M$ others,
its signal is split into $2^M$ peaks at frequencies of
$(\omega_0^n/\pi \pm J^{m_1n} \pm\cdots\pm J^{m_Mn})/2$,
one for each combination of ``up'' and ``down''
states for the $M$ spins to which it is coupled.
If the transverse magnetization is due to a $\pi/2$
rotation of a spin polarized along $\LAB{z}$ as before,
then the heights of these peaks are proportional to
the probability differences between pairs of states
$\KET{\kappa^{m_1}\ldots\kappa^n\ldots\kappa^{m_M}} \leftrightarrow
\KET{\kappa^{m_1}\ldots(1-\kappa^n)\ldots\kappa^{m_M}}$
separated by flips of that spin.
To show this, we restrict ourselves
to two spins for ease of presentation,
and consider a general diagonal
density operator of the form
\begin{equation} \begin{array}{rl}
\RHO_\LAB{zz} ~= & p_0 \KET{00}\BRA{00} + p_1 \KET{01}\BRA{01}
+ p_2 \KET{10}\BRA{10} + p_3 \KET{11}\BRA{11} \\
= & \FRAC14 + \HALF(p_0+p_1-p_2-p_3)\IZ^1 + \HALF(p_0-p_1+p_2-p_3)\IZ^2 \\
& \qquad +\, (p_0-p_1-p_2+p_3) \IZ^1\IZ^2 ~,
\end{array} \end{equation}
where $p_k$ denotes the probability
that a molecule is in the state $\KET{k}$.
Rotating this to
\begin{equation} \begin{array}{rcl}
\RHO_\LAB{xz} &\equiv& e^{-\imath\pi\IY^1}
\RHO_\LAB{zz} e^{\imath\pi\IY^1} \\
&=& \FRAC14 + \HALF(p_0+p_1-p_2-p_3)\IX^1
+ \HALF(p_0-p_1+p_2-p_3)\IZ^2 \\ && +\,
(p_0-p_1-p_2+p_3) \IX^1\IZ^2
\end{array} \end{equation}
and computing the signal in the Zeeman frame yields
\begin{equation} \begin{array}{rcl}
&& \TR\left( e^{-\imath t 2\pi J^{12} \IZ^1\IZ^2} \RHO_\LAB{xz} \,
e^{\imath t 2\pi J^{12} \IZ^1\IZ^2} (\IX^1 + \imath\IY^1) \right) \\
&=& \HALF \left( (p_0+p_1-p_2-p_3) \cos(\pi J^{12} t) \right. \\
&& +\, \left. (p_0-p_1-p_2+p_3) \, \imath \sin(\pi J^{12} t) \right) \\
&=& \HALF e^{\imath\pi J^{12}t} (p_0 - p_2)
+ \HALF e^{-\imath\pi J^{12}t} (p_1 - p_3) ~,
\end{array} \end{equation}
thus showing that the peaks at $\omega_0^1 \pm \pi J^{12}$ have
amplitudes proportional to the probability differences as claimed.

In closing, we mention that although vector
interpretations of single quantum inphase ($\IX^1$)
and antiphase ($\IX^1\IZ^2$) states are available
(and widely used in NMR \cite{Freeman:98}),
no satisfactory geometric interpretation
of general product states is known.
The development of an intuitive model
for the geometry determined by the action
of $\LAB{SU}(2^N)$ on the product operators
thus stands as an open problem in the field.
There are two reasons why the problem is nontrivial.
The first is the well-known the existence of
{\em correlated\/} states, whose density operators
cannot be factorized; in the case of a pure state,
these states are also called {\em entangled\/} \cite{Peres:95}.
We shall consider such states further in Section 5.
The second, much less widely recognized reason is
that there is but one imaginary unit for all the spins,
so that in the tensor product of their geometric algebras
the unit pseudo-scalars $8\IX^n\IY^n\IZ^n$ must be identified
by taking an appropriate quotient \cite{SomCorHav:98}.
This is a form of implicit correlation which is
always present even in otherwise factorizable states.
Further discussion of this issue may be found
in Refs.\ \cite{DorLasGul:93,DoLaGuSoCh:96}.

\vspace*{0.20in}
\section{Pseudo-pure state preparation and scaling}
Liquid state NMR must be done at temperatures
far above the differences between the spin
Hamiltonian's energy levels (eigenvalues).
The ensemble's spin state thus represents a
compromise between the constant force of the applied
magnetic field and the forces of the random fields induced
by the thermal motions of spins in other molecules.
Thus pure states are not available,
so that the underlying ensemble is not
uniquely determined by its density operator.
This would seem to make NMR useless as a means
of performing deterministic computations,
but in fact a class of mixed states has been found
for which a state vector is (up to an overall phase)
canonically associated with the density operator.
This section is devoted to describing the properties
and preparation of such {\em pseudo-pure states\/},
with emphasis on the computationally important issue
of how they scale with the number of spins $N$.

According to the principles of quantum
statistical mechanics \cite{Blum:96},
the density operator $\RHO_\LAB{eq}$ for an ensemble
of $N$-spin molecules at thermal equilibrium is
given by the Boltzman operator determined by their
common Hamiltonian, $\exp(-\VEC{H}/k_\LAB{B}T)$,
divided by the corresponding partition function
$Z_\LAB{eq} =$ \linebreak
$\TR(\exp(-\VEC{H}/k_\LAB{B}T))$
(where $k_B$ is Boltzman's constant).
The Hamiltonian $\VEC{H}$ is well-approximated
by its dominant Zeeman term $\VEC{H}_\LAB{Z}
= -\omega_0^1\IZ^1-\cdots-\omega_0^N\IZ^N$.
Given the gyromagnetic ratios of nuclear spins
and the strongest available magnetic fields, we have
$\|\VEC H_\LAB{Z}\| / (k_\LAB{B}T) \sim 10^{-5}$
at the temperatures needed for liquid-state NMR,
so that a linear approximation is quite accurate:
\begin{equation}
\RHO_\LAB{eq} ~\approx~ \frac{1 - \VEC H_\LAB{Z} / k_\LAB{B}T}
{\TR( 1 - \VEC H_\LAB{Z} / k_\LAB{B}T )} ~=~
\frac{1 - \VEC H_\LAB{Z} / k_\LAB{B}T}{2^N}
\end{equation}
In homonuclear (i.e.\ single spin isotope) systems,
one can assume that $\omega_0^n \equiv \hbar B_0 (1 - \sigma^n)
\gamma^n \approx \hbar B_0 \gamma$ is constant for all $n$.
Since the amplitude of an NMR signal is also
proportional to imprecisely known factors
determined by the spectrometer setup,
$\omega_0^n/(k_\LAB{B}T)$ is usually set to
unity when analyzing a homonuclear experiment
(or to the ratios of each $\gamma^n$
with $\FUN{min}_m(\gamma^m)$ otherwise).
The partition function $2^{-N}$ is likewise
constant for any given system, but because
of our interest in scaling we shall always
include it explicitly in this section.

It is important to observe that,
because the angular momentum components observed
by NMR have no scalar part (i.e.\ are traceless),
the scalar part (identity component) of the density
operator $2^{-N}$ does not contribute to the signal.
It also does not evolve under unitary transformations,
and hence NMR spectroscopists
usually forget about it altogether ---
even though it comprises the vast majority
of the norm of the density operator.
In these terms, the equilibrium density operator
of a two-spin system, and its matrix representation
in the usual computational basis, is
\renewcommand{\arraystretch}{1.0}%
\begin{equation} \begin{array}{rcl}
&& \hat{\RHO}_\LAB{eq} ~=~ \FRAC{1}{4} (\IZ^1 + \IZ^2)
~=~ \FRAC{1}{4} (\KET{00}\BRA{00} - \KET{11}\BRA{11})
\vspace*{3pt} \\ &~\leftrightarrow~&
\FRAC{1}{4}\, \MAT{Diag}( 1, 0, 0, -1 )
~\equiv~ {\displaystyle\frac{1}{4}}
\left[ \begin{array}{rrrr}
1 & 0 & 0 & 0 \\ 0 & 0 & 0 & 0 \\
0 & 0 & 0 & 0 \\ 0 & ~~0 & ~~0 & -1
\end{array} \right] ~,
\end{array} \end{equation}
\renewcommand{\arraystretch}{1.3}%
where the ``hat'' on $\RHO_\LAB{eq}$
signifies its traceless part.

In contrast, the density operator of two spins in their
pseudo-pure ground (assuming $\gamma > 0$ as usual) state is
\begin{equation} \begin{array}{rcl}
\hat{\RHO}_{00} &\equiv& \pm\FRAC{1}{6}
(\IZ^1 + \IZ^2 + 2 \IZ^1 \IZ^2) ~=~
\pm\FRAC{1}{3} \left( \EP^1 \EP^2 - \FRAC{1}{4} \right)
\\ &=&
\pm\FRAC{1}{3} \left( \KET{00}\BRA{00} - \FRAC{1}{4} \right)
~\leftrightarrow~ \pm\FRAC{1}{12}\, \MAT{Diag}( 3, -1, -1, -1 ) ~.
\end{array} \end{equation}
The overall sign depends on whether we have a
population excess or deficit in the ground state;
for consistency, we shall generally assume the former.
Observe that a unitary transformation of the density
operator induces a transformation of the corresponding
state vector just as it does for true pure states, since
\begin{equation} \label{eq:pp_tfm}
\VEC{U} \hat{\RHO}_{00}\, \tilde\VEC{U} ~=~
\FRAC{1}{3} \left( \left( \VEC{U} \KET{00} \right)
\left( \VEC{U} \KET{00} \right)^{\sim} - \FRAC{1}{4} \right) ~.
\end{equation}
Similarly, because the NMR observables $\VEC A = \IX^n, \IY^n$
are traceless, the ensemble-average expectation value relative
to a pseudo-pure density operator yields the ordinary
expectation value versus the corresponding state vector:
\begin{equation} \label{eq:pp_obs}
\TR( \VEC A\, \hat{\RHO}_{00} ) ~=~ \FRAC{1}{3} \left( \TR
( \VEC{A} \KET{00}\BRA{00} ) - \FRAC{1}{4} \TR( \VEC{A} ) \right)
~=~ \FRAC{1}{3} \BRA{00} \VEC{A} \KET{00}
\end{equation}
The general form of a pseudo-pure density operator is
\begin{equation} \label{eq:ppform}
\hat{\RHO}_\LAB{\psi} ~=~ \FRAC{N/2}{\rule{0pt}{6pt}2^N-1}
\left( \KET{\psi}\BRA{\psi} - 2^{-N} \right) ~,
\end{equation}
where $\KET{\psi}$ is a normalized $N$-spin state vector,
and the prefactor has been chosen so as to keep
the maximum eigenvalue $\| \hat{\RHO}_{\psi} \|_2$ equal
to that of the $N$-spin equilibrium density operator.

Even though we have defined them
to have the same maximum eigenvalue,
for $N > 1$ the remaining eigenvalues of
$\hat{\RHO}_\LAB{eq}$ and $\hat{\RHO}_{\psi}$ differ,
and hence there is no unitary
transformation taking one to the other.
There are nevertheless a number
of nonunitary processes by which
one can prepare pseudo-pure states.
The most direct is to generate a spatially
varying distribution of states across the sample,
such that the ensemble average is pseudo-pure.
This can be done by using a {\em field gradient\/}
along the $\LAB{z}$-axis to create a
position-dependent phase shift whose average is zero,
thereby in effect setting the transverse ($\LAB{xy}$)
components of the density operator to zero.\footnote{
In the homonuclear case, the zero-quantum coherences
are not rapidly dephased by a $\LAB{z}$-gradient,
so a slightly more complicated procedure is necessary. }
For example, it is readily shown that the sequence
\begin{equation}
[\FRAC{\pi}{4}(\IX^1+\IX^2)] \rightarrow [\pi\IZ^1\IZ^2]
\rightarrow [-\FRAC{\pi}{6}(\IY^1+\IY^2)]
\end{equation}
applied to the two-spin equilibrium state $\hat{\RHO}_\LAB{eq}$ yields
\begin{equation}
2^{-\frac52} \left( \sqrt3\, \left( \EP^1\EP^2 - \FRAC14 -
\IX^1\IX^2 \right) - \IX^1\EM^2 - \EM^1\IX^2 \right) ~,
\end{equation}
which is reduced by a $\LAB{z}$-gradient
to $(3/32)^{\frac12}(\EP^1\EP^2 - 1/4)$.
Further RF and gradient pulse sequences
which convert the equilibrium state of
two and three spin systems to pseudo-pure states
may be found in Ref.~\cite{CorPriHav:98}.

An alterative proposed by E.~Knill {\em et al.\/}
\cite{KniChuLaf:98} is to ``time-average''
the results of several separate experiments.
In the simple case of a two spin system,
the average of the three states
\begin{equation} \begin{array}{rcl}
\hat{\RHO}_{123} ~~\equiv~& \FRAC{1}{4} (\IZ^1 + \IZ^2)
&\leftrightarrow~~ \FRAC{1}{4}\, \MAT{Diag}( 1, 0, 0, -1 ) \\ 
\hat{\RHO}_{231} ~~\equiv~& \FRAC{1}{4} (\IZ^1 + 2 \IZ^1 \IZ^2)
&\leftrightarrow~~ \FRAC{1}{4}\, \MAT{Diag}( 1, 0, -1, 0 ) \\ 
\hat{\RHO}_{312} ~~\equiv~& \FRAC{1}{4} (2 \IZ^1 \IZ^2 + \IZ^2)
&\leftrightarrow~~ \FRAC{1}{4}\, \MAT{Diag}( 1, -1, 0, 0 )
\end{array} \end{equation}
is the pseudo-pure state
\begin{equation} \begin{array}{rcl} \label{eq:cyclic_avg}
\FRAC{1}{3} (\hat{\RHO}_{123} + \hat{\RHO}_{231} + \hat{\RHO}_{312})
&=& \FRAC{1}{12} (2 \IZ^1 + 2 \IZ^2 + 4 \IZ^1 \IZ^2)
\\ &~\leftrightarrow~&
\FRAC{1}{12}\, \MAT{Diag}( 3, -1, -1, -1 ) ~.
\end{array} \end{equation}
More generally, one can obtain the same
results that one would get on a pseudo-pure
state by averaging the results of the experiments
over all $2^N-1$ cyclic permutations of the nonground
state populations of the equilibrium density operator.
Although this naive approach is not efficient,
Knill {\em et al.\/} have shown that one can average over
smaller groups in time $O(N^3)$ with much the same effect.

A fundamentally different approach, first proposed
by Stoll, Vega \& Vaughan \cite{StoVegVau:77}
and subsequently adapted to NMR computing by
Gershenfeld \& Chuang \cite{GershChuan:97},
involves working with subpopulations of molecules
distinguished by the states of additional {\em ancilla\/} spins.
Gershenfeld and Chuang \cite{ChGeKuLe:98} have given an
example of a two-spin {\em conditional pseudo-pure state\/}
(as we call it), which is obtained by row/column
permutation of the diagonal equilibrium density matrix
$\MAT{Diag}( 3, 1, 1, -1, 1, -1, -1, -3 ) / 16$
of a three-spin system including one ancilla, namely
\begin{equation} \begin{array}{rcl} \label{eq:condpure3a}
&&\FRAC{1}{16}\, \MAT{Diag}( 3, -1, -1, -1, -3, 1, 1, 1 )
\\ &~\leftrightarrow~&
\FRAC{1}{16} (2\IZ^1 (2\IZ^2 + 2\IZ^3 + 4\IZ^2\IZ^3))
\\ &=&
\FRAC{1}{4} (\EP^1 - \EM^1) (\EP^2\EP^3 - \FRAC{1}{4}) ~.
\end{array} \end{equation}
The last form makes it clear that in
the subpopulation with the first spin ``up'',
which is labeled by $\EP^1$,
and in the subpopulation with it ``down'',
which is labeled by $\EM^1$,
spins $2$ and $3$ are in the
pseudo-pure state $\EP^2\EP^3 - 1/4$.
Since the spectrum of spins $2$ and $3$ is
antiphase with respect to the ancilla spin $1$,
one can select the subpopulations just by
keeping only either positive or negative peaks.
Although the situation is considerably
more complicated with more spins,
Gershenfeld and Chuang have shown that conditional
pure states can be obtained (with some loss of signal)
using as few as $O(\log(N))$ ancillae.

An alternative to conditional pure states,
which we call {\em relative pseudo-pure states\/},
can be obtained via the partial trace operation (in NMR,
decoupling \cite{ErnBodWok:87,Freeman:98,Munowitz:88,Slichter:90}),
rather than peak selection as above.
For example, a two-spin relative pseudo-pure state is given
by the partial trace over the ancilla spins 1 \& 2 in
\begin{equation} \begin{array}{rcl} \label{eq:relpure4}
&& \FRAC{1}{32}\, \MAT{Diag}( 4, 2, 2, 0, 2, 0,
-2, 0, 0, -2, 0, 2, 0, -2, -2, -4 )
\\ &\leftrightarrow& \FRAC{1}{16}
\left( \EP^1\EP^2 (\EP^3 + \EP^4) +
\EP^1\EM^2 (\EP^3\EP^4 - \EM^3\EP^4)
\right. \\ &&  ~+ \left.
\EM^1\EP^2 (\EM^3\EM^4 - \EP^3\EM^4) -
\EM^1\EM^2 (\EM^3 + \EM^4) \right) ~,
\end{array} \end{equation}
which is again a permutation of the
diagonal elements of $\hat{\RHO}_\LAB{eq}$.
This can be seen by adding up the $4\times 4$ blocks of
the matrix, obtaining $\MAT{Diag}( 6, -2, -2, -2 ) / 32$.
Alternatively, since the partial trace in the product
operator formalism corresponds to simply eliminating
those terms depending on either spins $1$ or $2$ and
multiplying the remaining terms by $4$ \cite{SomCorHav:98},
we need only add up the multipliers of
$\EP^1\EP^2, \ldots, \EM^1\EM^2$, which yields
\begin{equation} \begin{array}{rcl}
&& \FRAC{1}{16} ((1 + \EP^3 - \EM^3) (1 + \EP^4 - \EM^4) - 1)
\\ &=&
\FRAC{1}{4} \left( \EP^3 \EP^4 - \FRAC{1}{4} \right)
~\leftrightarrow~
\FRAC{1}{32}\,\MAT{Diag}( 6, -2, -2, -2 ) ~.
\end{array} \end{equation}

We now consider briefly how
the SNR (signal-to-noise ratio)
of these methods of creating pseudo-pure
states scales with the number of spins $N$.
It has been argued that since the equilibrium
population of the ground state falls off
exponentially with the number of spins,
and all these methods are aimed in some fashion at
isolating the signal from the ground state population,
the SNR of all these methods must likewise
decline exponentially with $N$ \cite{Warren:97}.
Although this argument carries considerable weight,
we shall see that the number and variety of the available
methods renders the actual situation rather more complex.
The standard to which the signal strength must be compared
is that of a single spin in its equilibrium state, namely
\begin{equation}
\hat{\RHO}_\LAB{eq} ~=~ \HALF \IZ^1 ~=~ \FRAC{1}{4}
(\KET{0}\BRA{0} - \KET{1}\BRA{1}) ~.
\end{equation}
The maximum eigenvalue $\| \hat{\RHO}_\LAB{eq} \|_2 = 1/4$
is what we will use as the standard signal strength for
spins of like gyromagnetic ratio (as assumed throughout).
We shall therefore calculate the SNR of a pseudo-pure
state by transforming it to the corresponding ground state
$\KET{0\cdots 0}\BRA{0\cdots 0} - 2^{-N}$ (if need be),
taking the partial trace over all but one of the spins,
and multiplying the maximum eigenvalue of the result by $4$.

\begin{figure}[t]
\begin{picture}(300,250)
\put(5,15){ \psfig{file=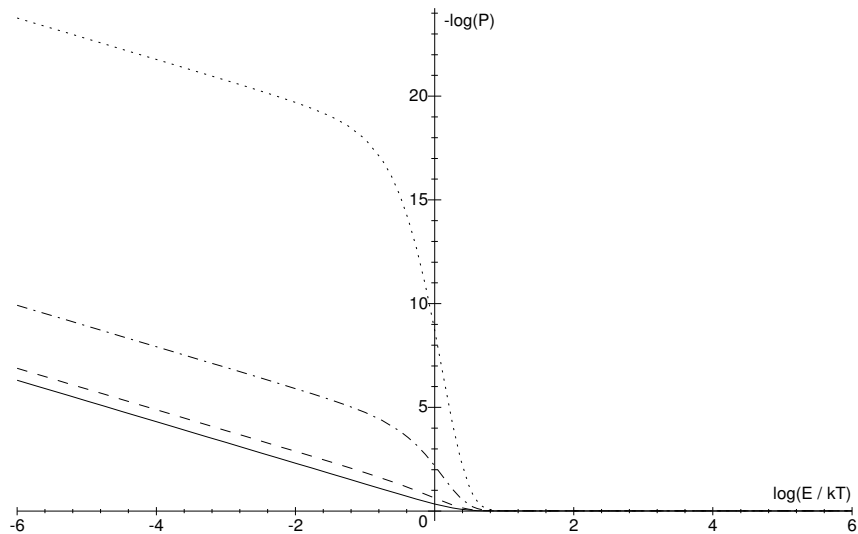,width=4.5in}
} \end{picture} \caption{
Negative base ten logarithm of the polarization
$P$ as a function of the logarithm of the ratio
of the energy level spacing to $k_\LAB{B}T$ for
a one (solid), four (dash), sixteen (dash-dot) and sixty-four (dot)
spin pseudo-pure state obtained by cyclic averaging.
For protons in a standard 500 MHz spectrometer at room temperature,
$E / (k_\LAB{B}T) \approx P \sim 10^{-5}$ at equilibrium. }
\label{fig:polar}
\end{figure}

The maximum eigenvalue of the partial trace over all
but one of the spins in a pseudo-pure state obtained
by cyclic averaging, as in Eq.~(\ref{eq:cyclic_avg}),
is easily seen to be $N/(4(2^N-1))$, which decays
almost exponentially with the number of spins $N$.
There is an additional factor of $\sqrt{2^N-1}$
which comes from averaging over $2^N-1$ experiments,
and gives a net SNR of $N/(4\sqrt{2^N-1})$ for the average.
The exponential time requirements of cyclic averaging
will nonetheless force one to average over smaller groups,
with consequently smaller improvements in the SNR.
In any case, the SNR declines superpolynomially with $N$.
Figure \ref{fig:polar} shows how the signal strength changes as
a function of the ratio of the energy level spacing to $k_\LAB{B}T$,
relative to the signal in a perfectly polarized sample,
when the pseudo-pure state is obtained by cyclic averaging,
for varying numbers of spins.

Because of the many possible variations on the
ideas and the difficulty of analyzing all of them,
it is not practical to present simple formulae for the
SNR of the other methods of preparing pseudo-pure states.
Further complexity is added to the situation by
the ability to combine the various methods above.
A number of such combinations are given in Knill
{\em et al.\/} \cite{KniChuLaf:98},
along with bounds on the SNR for each.
In our laboratory we are developing a new
method, again based on field gradients,
which enables the sample to be divided into
discrete volumes and separate unitary
transformations to be applied to each.
In principle, this permits multiple
experiments to be performed, and their
results added, in a single experiment,
thereby performing an average over
multiple experiments in constant time.
This new method could also be used in a
variety of combinations with existing methods.

It is nevertheless encouraging to observe that
the SNR of the two-spin conditional and relative
pseudo-pure states given in Eqs.~(\ref{eq:condpure3a})
and (\ref{eq:relpure4}) is $1/2$ in both cases;
this is exactly the decline in the ground state
population of a two-spin system compared to a one-spin.
In Eq.~(\ref{eq:condpure3a}), we attain this ``theoretical limit''
because the expansion of the density operator consists of
a single term conditioned on the state of a single ancilla;
it is not possible to do as well with more spins.
In Eq.~(\ref{eq:relpure4}), however,
it is because such permutations are able to
concentrate polarization in a subset of the spins.
We have found this makes it possible to derive
a two-spin pseudo-pure state from a six-spin
equilibrium state with {\em no\/} loss of SNR,
whereas a simplistic ground-state population
argument implies we should lose at least $1/2$.
This may be seen by adding up the rows in
the rearrangement of $\hat{\RHO}_\LAB{eq}$
shown in Eq.~(\ref{eq:relpure6}) below,
which corresponds taking the traces of the
four $16\times 16$ blocks along the diagonal,
and yields $\MAT{Diag}( 48, -16, -16, -16 )$.
\renewcommand{\arraystretch}{1.1}%
{%\addtolength{\arraycolsep}{-4pt}
\begin{equation} \label{eq:relpure6}
\MAT{Diag} \begin{array}[t]{rrrrrrrrrrrrrrrrr} (
~6, &  4, &  4, &  4, &  4, &  4, &  4, &  2, & 
 2, &  2, &  2, &  2, &  2, &  2, &  2, &  2, & \\
 0, &  0, &  0, &  0, &  0, &  0, &  0, &  0, & 
 0, &  0, & -2, & -2, & -2, & -2, & -4, & -4, & \\
 0, &  0, &  0, &  0, &  0, &  0, &  0, &  0, & 
 0, &  0, & -2, & -2, & -2, & -2, & -4, & -4, & \\
~2, & ~2, & ~2, & ~2, & ~2, & ~2, & -2, & -2, & 
-2, & -2, & -2, & -2, & -2, & -4, & -4, & -6\,  &
~) \end{array}
\end{equation}
}%
\renewcommand{\arraystretch}{1.3}%
The partial trace over one of the two remaining
spins then gives $\MAT{Diag}( 32, -32 )$,
which when divided by $128$ (twice the partition
function) yields $\FRAC{1}{2} \IZ$ as claimed.

A general algorithm has recently been given by
Schulman \& Vazirani \cite{SchulVazir:98} whereby one
can ``distill'' an $M$-spin relative pure state from
an ensemble of molecules each containing $N$ spins.
Starting from a uniform polarization of $P$,
this algorithm yields $M$ perfectly
polarized spins providing $M/N \sim O(P^2)$,
a result anticipated by earlier work in NMR which showed
that the polarization of a single spin can be enhanced
by at most a factor $O(\sqrt{N})$ \cite{Sorensen:89}.
Unfortunately, given that $P \approx 10^{-5}$ for
protons at equilibrium in a standard 500 MHz spectrometer,
a molecule with of order $10^{10}$ spins would be needed
to prepare a perfectly polarized state on a single spin
--- which is in a pseudo-pure state at equilibrium!
The importance of Schulman and Vazirani's algorithm
thus lies in the fact that it shows that there
is sufficient order in a typical NMR sample of
$10^{20}$ spins at room temperatures to make
it at least theoretically possible to perform
quantum computations on of order $10^{10}$ spins.

One might hope that a more tractable algorithm,
in terms of the absolute resources required,
could be found by requiring only that it produce
a {\em pseudo\/}-pure state with bounded SNR
from the high-temperature equilibrium state.
Since in the high-temperature approximation
the largest element of the density matrix decays
exponentially with the number of spins, it is clear that
any such an algorithm must go beyond that approximation.
Even so, given that Schulman and Vazirani's algorithm is
currently far beyond our ability to implement physically,
it seems unlikely that a practical breakthrough
will be obtained by purely algorithmic means.
Fortunately, physical methods of ``refrigerating'' the spins
are available, for example optical pumping \cite{NSRATP:96}.
These are presently confined to very simple systems,
but such a source of polarization could in principle be
used in conjunction with polarization transfer techniques
to produce (pseudo$\mbox{-}$)pure states on large numbers of spins.
Even at the low polarizations we can conveniently access,
however, NMR has proved itself to be a powerful means of
exploring quantum dynamics in Hilbert spaces of substantial size.
To illustrate this, we will now present the results of NMR
experiments which constitute a macroscopic analogue of a
quantum mechanical test for quantum correlations that are
inconsistent with the existence of ``hidden variables''.

\vspace*{0.20in}
\section{Macroscopic analogues of quantum correlations}
Given the success of the purely classical Bloch equations
(and their multispin extensions) in describing liquid-state NMR
phenomena \cite{ErnBodWok:87,Freeman:98,Munowitz:88,Slichter:90},
it is perhaps surprising that experiments can be
performed whose mathematical description, at least,
is formally identical to that of experiments which are
believed to demonstrate uniquely ``quantum'' phenomena.
For example, Seth Lloyd has recently proposed that
the nonclassical correlations in (Mermin's version of)
the GHZ state can be validated using NMR \cite{Lloyd:98}.
His approach involves using a fourth ``observer'' spin
to perform a nondemolition measurement on the three
spins in a GHZ state (or a pseudo-pure analogue thereof).
Here we shall describe experiments which demonstrate
another, rather different way in which we can
``emulate'' quantum phenomena with liquid-state NMR.
In reading this account, it should be kept in mind
that although pseudo-pure states do provide a faithful
representation of the transformations of pure states
within the highly mixed states that are available
in liquid-state NMR, their physical interpretation
differs significantly from that of true pure states.
Hence, as discussed in greater detail at the end of
this section (and also in Ref.\ \cite{BraunsteinEtAl:98}),
our results should {\em not\/} be taken to resolve any
foundational issues in quantum mechanics \cite{Peres:95}.
They demonstrate, nonetheless, a degree of coherent
control sufficient to enable such issues to be addressed,
{\em if\/} these same transformations and measurements
were applied to a true pure state.

The approach taken here was inspired by an educational
paper published a few years ago, in which T.~F.~Jordan
has shown that the contradictions with hidden variables
implied by violations of Bell's inequalities as well as
by the GHZ and Hardy's paradox can be derived entirely by
consideration of the expectation values of product operators,
rather than by observations on single spins \cite{Jordan:94b}.
This shows that, in principle, it is not necessary to use
nondemolition measurements with an observer spin in order to
perform experiments which demonstrate these contradictions by NMR;
it can be done directly from observations on ensembles of the spins
of interest, providing at least they are in a (pseudo-)pure state.
In a companion paper to Jordan's, N.~D.~Mermin points out that in
real-life experiments it is nevertheless not possible to perform
the measurements, either of single spins nor (by implication)
of expectation values, with sufficient precision to establish the
``perfect'' (total) correlations on which ``EPR'' arguments against
the existence of hidden variables are based \cite{Mermin:94}.
In that same paper, however, Mermin shows that Hardy's paradox
is a special case of the Clauser-Horne form of Bell's inequality.
This enables Hardy's paradox \cite{Branning:97,Hardy:93} to be
extended to an open set in the Hilbert space of only two spins,
to which sufficiently precise experimental data can confine us.

In the following, we present the results of NMR experiments
which implement the specific example of Hardy's paradox
presented by Mermin in an Appendix to his paper \cite{Mermin:94}.
Let us map the ``red'' and ``green'' eigenstates $\KET{\LAB{1G}}$
and $\KET{\LAB{1R}}$ of Mermin's measurement $1$ to
the spin states $\KET{0}$ and $\KET{1}$, respectively.
It will be clearer here to relabel this measurement as ``$\LAB{A}$'',
and to use $\KET{\alpha_\LAB{G}} \equiv \KET{0}$ and
$\KET{\alpha_\LAB{R}} \equiv \KET{1}$ as synonyms for its eigenbasis.
Correspondingly, we will relabel Mermin's measurement
$2$ as ``$\LAB{B}$'', and denote its the eigenbasis by
\begin{equation}
\KET{\beta_\LAB{G}} ~\equiv~ \SQRT{\FRAC{3}{5}} \KET{0}
- \SQRT{\FRAC{2}{5}} \KET{1} \quad\mbox{and}\quad
\KET{\beta_\LAB{R}} ~\equiv~ \SQRT{\FRAC{2}{5}} \KET{0}
+ \SQRT{\FRAC{3}{5}} \KET{1} ~.
\end{equation}
Then the state which Mermin has shown leads to a
near-maximum violation of Bell's inequality while
also providing an example of Hardy's paradox is
\begin{equation} \label{eq:sigh}
\KET{\psi} ~\equiv~ \HALF \KET{00} + \SQRT{\FRAC{3}{8}}
\KET{01} + \SQRT{\FRAC{3}{8}} \KET{10} ~.
\end{equation}

To translate this into the context of NMR,
we first note that the observable whose expectation
value is the probability that measurement $\LAB{A}$
yields the state $\KET{\alpha_\LAB{G}}$ is
given by $\VEC A \equiv \EP = \HALF(1+2\IZ)$
(we drop the usual spin index because the measurements
$\LAB{A}$ \& $\LAB{B}$ are assumed the same for both spins).
Similarly the observable which gives the probability that
measurement $\LAB{B}$ yields $\KET{\beta_\LAB{G}}$ is
\begin{equation} \begin{array}{rcl}
\VEC B ~\equiv~ \KET{\beta_\LAB{G}}\BRA{\beta_\LAB{G}}
&=& \FRAC{3}{5} \KET{0}\BRA{0} + \FRAC{2}{5} \KET{1}\BRA{1} -
\SQRT{\FRAC{6}{25}} (\KET{0}\BRA{1} + \KET{1}\BRA{0}) \\ 
&=& \HALF + \FRAC{1}{5} \IZ - \SQRT{\FRAC{24}{25}} \IX ~.
\end{array} \end{equation}
In addition, the density operator
(including the identity) of Mermin's state is
\begin{equation} \begin{array}{rcl} \label{eq:Sigh}
\EMB\Psi ~\equiv~ \KET{\psi}\BRA{\psi}
&=& \FRAC{1}{4} + \FRAC{1}{8} \left(\IZ^1 + \IZ^2\right)
- \HALF \IZ^1\IZ^2 \\ 
&& +\, \SQRT{\FRAC{3}{8}} \left(\IX^1(1 + 2\IZ^2) +
(1 + 2\IZ^1)\IX^2\right) \\ 
&& +\, \FRAC{3}{4}\left( \IX^1\IX^2 + \IY^1\IY^2 \right) ~.
\end{array} \end{equation}

The state $\KET{01}$ is obviously related
to $\KET{00}$ by a rotation of spin $2$,
while $\KET{00}$ can likewise be rotated to
$\KET{10}$, but without affecting $\KET{01}$,
by a {\em conditional\/} rotation of spin $1$.
We shall denote these by
\begin{equation}
\VEC P(\phi) ~\equiv~ e^{-\imath\phi\IY^2}
\quad\mbox{and}\quad
\VEC Q(\theta) ~\equiv~ e^{-\imath\theta\IY^1\EP^2} ~,
\end{equation}
respectively.
They act consecutively on the ground state to yield
\begin{equation}
\BRA{00} \tilde{\VEC P}(\phi) \tilde{\VEC Q}(\theta) ~=~
\left[ \,\cos(\theta/2) \cos(\phi/2),\, \sin(\theta/2)
\cos(\phi/2),\, \sin(\phi/2),\, 0\, \right] ~,
\end{equation}
which is easily verified to equal %the desired vector
$\BRA{\psi} = [ 1/2, \sqrt{3/8}, \sqrt{3/8},\, 0 ]$ when
\begin{equation}
\phi ~=~ 2 \arctan( \SQRT{3/5} )
\quad\mbox{and}\quad
\theta ~=~ 2 \arctan( \SQRT{3/2} ) ~.
\end{equation}
Using the product operator techniques presented in section 3,
these transformations are readily implemented by NMR pulse sequences.

The next thing to notice is that if we take expectation values
with the usual idempotents $\EP^1\EP^2, \ldots, \EM^1\EM^2$, we get
\begin{equation} \begin{array}{rcl}
\FRAC{1}{4} ~= & 4\left\langle \EMB\Psi \EP^1\EP^2 \right\rangle
& \equiv~ 4\left\langle \EMB\Psi \VEC A^1 \VEC A^2 \right\rangle
\\ 
\FRAC{3}{8} ~= & 4\left\langle \EMB\Psi \EP^1\EM^2 \right\rangle
& \equiv~ 4\left\langle \EMB\Psi \VEC A^1 (1 - \VEC A^2) \right\rangle
\\ 
\FRAC{3}{8} ~= & 4\left\langle \EMB\Psi \EM^1\EP^2 \right\rangle
& \equiv~ 4\left\langle \EMB\Psi (1 - \VEC A^1) \VEC A^2 \right\rangle
\\ 
0 ~= & 4\left\langle \EMB\Psi \EM^1\EM^2 \right\rangle
& \equiv~ 4\left\langle \EMB\Psi (1 - \VEC A^1) (1 - \VEC A^2)
\right\rangle ~.
\end{array} \end{equation}
These correspond to the diagonal of the
density matrix in the usual $\IZ$ basis,
\begin{equation}
{\bf diag}(\EMB\Psi) ~=~
[ \FRAC{1}{4}, \FRAC{3}{8}, \FRAC{3}{8}, 0 ]
\qquad\mbox{($\LAB{A}$ on 1, $\LAB{A}$ on 2)} ~,
\end{equation}
which contains the probabilities of the four possible outcomes
of performing measurement $\LAB{A}$ on both spins (as shown).

The product operator form of $\VEC B$ immediately
makes clear that measurement $\LAB{B}$ is just
a measurement of the magnetization of the spin
along an axis inclined at an angle of $\zeta
\equiv \arctan(\sqrt{24}) = \pi - \theta$ to
the $\LAB z$-axis in the $\LAB{xz}$-plane.
Letting $\VEC R \equiv \exp(-\imath \zeta \IY)$,
it follows that the probability that measurement $\LAB{B}$ on
spin $1$ yields ``$\LAB{G}$'' (i.e.~$\KET{\beta_\LAB{G}}$) is
\newcommand{\tildeR}% define \tilde{\VEC R} the height of \VEC{R}
{{\rule[0pt]{0pt}{7pt}\smash{\tilde\VEC{R}}}}
\begin{equation}
4 \AVG{ \EMB\Psi \VEC B^1 } ~=~
4 \AVG{ \EMB\Psi \tildeR^1
\VEC A^1 \VEC{R}^1 } ~=~
4 \AVG{ \VEC{R}^1 \EMB\Psi \tildeR^1 \VEC A^1 } ~.
\end{equation}
with a similar expression for spin $2$.
More generally, the probabilities of the outcomes
of the other combinations of measurements are given
by the diagonals of the transformed density matrices:
\begin{equation} \begin{array}{rcl}
{\bf diag}\left( \VEC{R}^2 \EMB\Psi \tildeR^2 \right)
&=& [ 0, \FRAC{5}{8}, \FRAC{9}{40}, \FRAC{3}{20} ]
\qquad\qquad\mbox{($\LAB{A}$ on 1, $\LAB{B}$ on 2)} \\
{\bf diag}\left( \VEC{R}^1 \EMB\Psi \tildeR^1 \right)
&=& [ 0, \FRAC{9}{40}, \FRAC{5}{8}, \FRAC{3}{20} ]
\qquad\qquad\mbox{($\LAB{B}$ on 1, $\LAB{A}$ on 2)} \\
{\bf diag}\left( \VEC{R}^1\VEC{R}^2
\EMB\Psi \tildeR^2 \tildeR^1 \right)
&=& [ \FRAC{9}{100}, \FRAC{27}{200}, \FRAC{27}{200}, \FRAC{16}{25} ]
\qquad\mbox{($\LAB{B}$ on 1, $\LAB{B}$ on 2)}
\end{array} \end{equation}
For compactness, let us denote these probabilities by $\Psi_{kl}^{ij}$,
where $i,j \in \{\LAB{A},\LAB{B}\}$ are the measurements and
$k,l \in \{\LAB{G},\LAB{R}\}$ are the corresponding outcomes,
e.g.~$\Psi_\LAB{GR}^\LAB{AB} = 4\AVG{\EMB\Psi \VEC A^1(1 - \VEC B^2)}$.

We may translate Mermin's proof \cite{Mermin:94}
that these probabilities are incompatible
with hidden variables associated with the
individual spins into this context as follows:
First, since $\Psi_\LAB{GG}^\LAB{AB} = \Psi_\LAB{GG}^\LAB{BA} = 0$,
in any molecule wherein one of the spins is parallel
to the $z$-axis the other must be antiparallel to
the axis of measurement $\LAB{B}$ and vice versa.
Hence, since $\Psi_\LAB{GG}^\LAB{BB}$ is nonzero,
in some molecules ($9$\%, to be precise) both
spins must be antiparallel to the $z$-axis.
But this contradicts the fact that $\Psi_\LAB{RR}^\LAB{AA} = 0$.
More generally, Mermin has shown that
\begin{equation}
\Psi_\LAB{GG}^\LAB{BB} ~\le~ \Psi_\LAB{GG}^\LAB{AB}
+ \Psi_\LAB{RR}^\LAB{AA} + \Psi_\LAB{GG}^\LAB{BA}
\end{equation}
is an example of the Clauser-Horne form of Bell's inequality \cite{Peres:95}.
Hence to disprove the existence of such one-particle hidden variables it
would be sufficient to determine these probabilities to $\pm2$\% or so.

\begin{figure}[t]
\begin{picture}(324,275)
\put(0,144){ \psfig{file=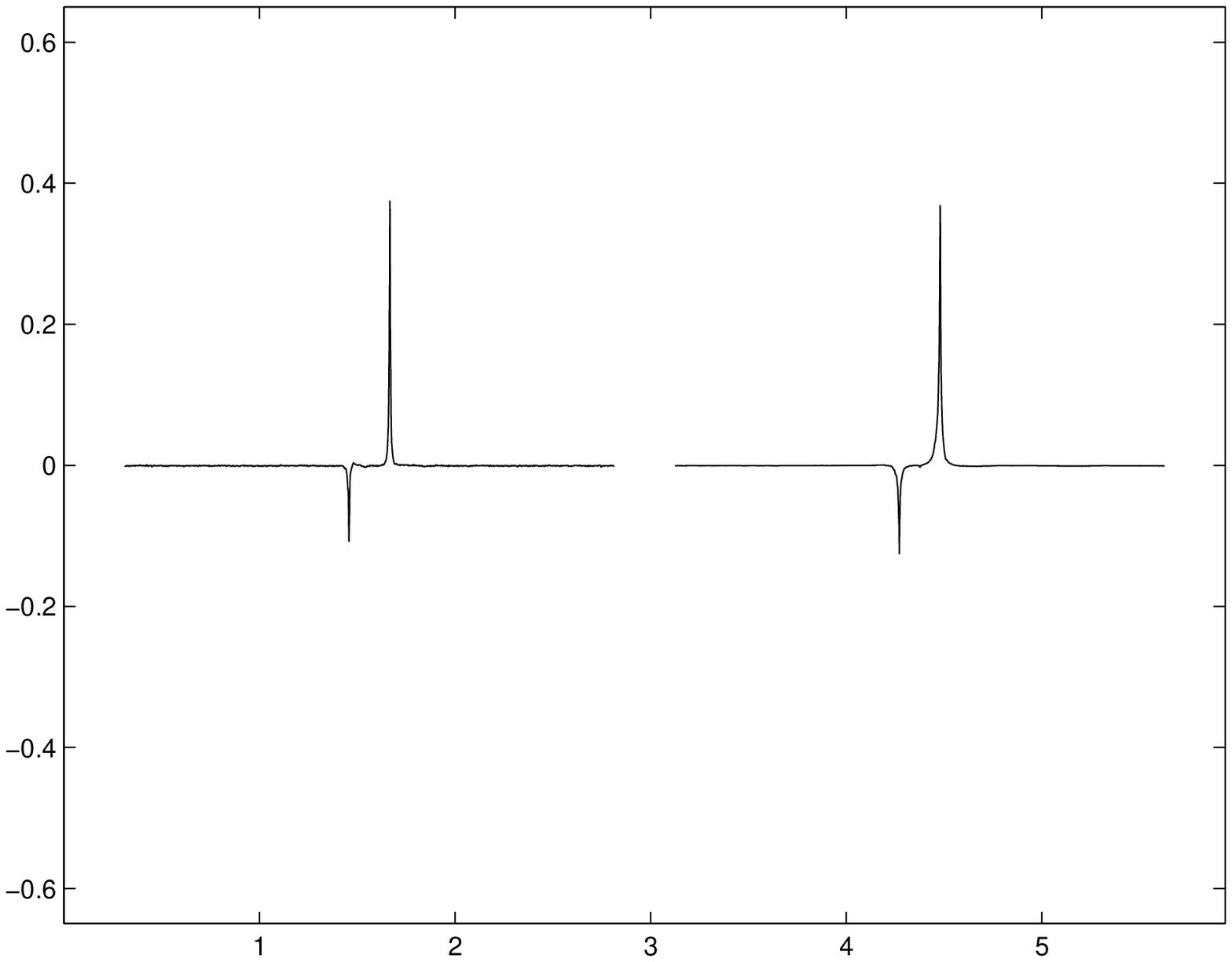,width=2.25in} }
\put(75,255){ $\LAB{A^CA^H}$ }
\put(170,144){ \psfig{file=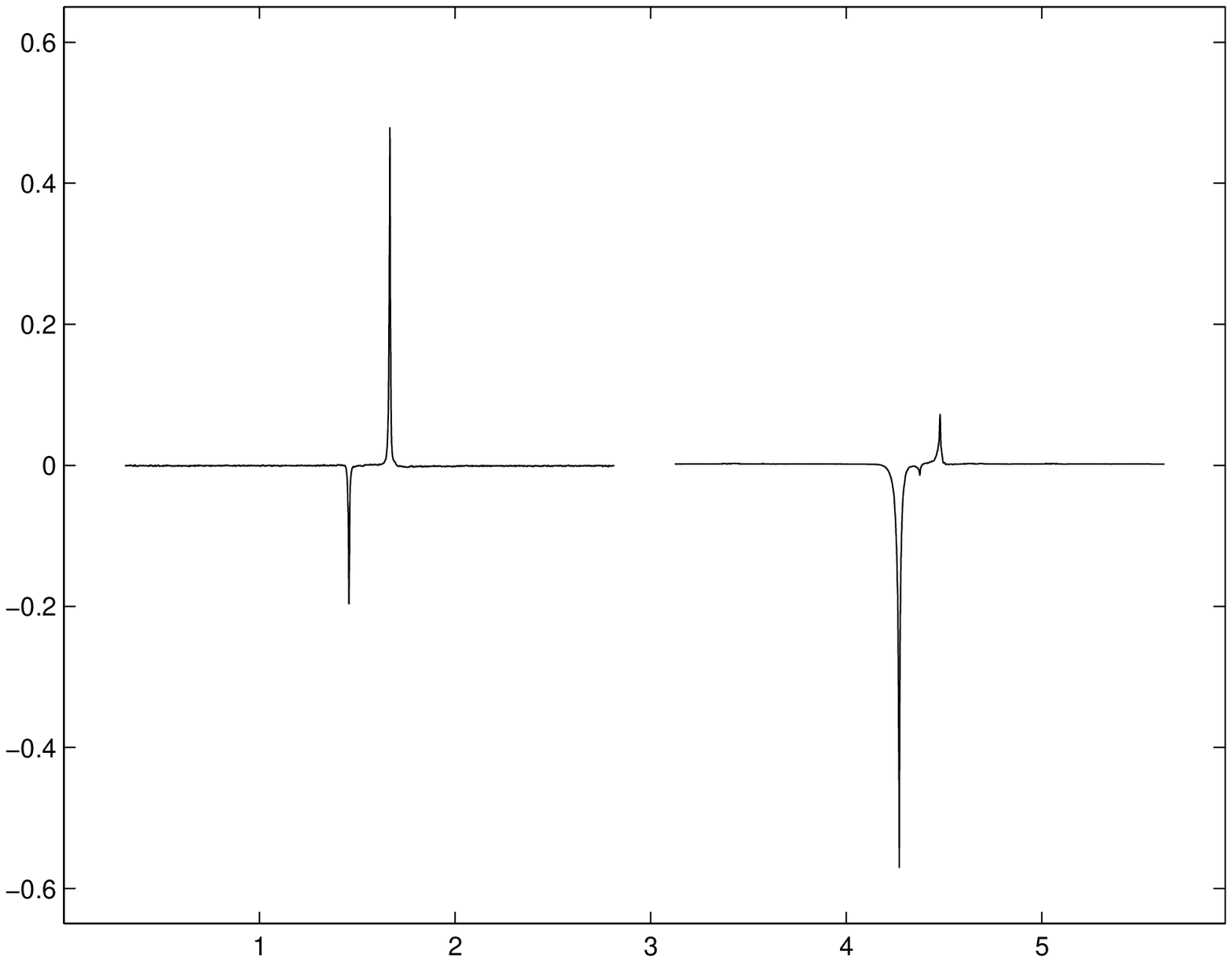,width=2.25in} }
\put(245,255){ $\LAB{A^CB^H}$ }
\put(0,5){ \psfig{file=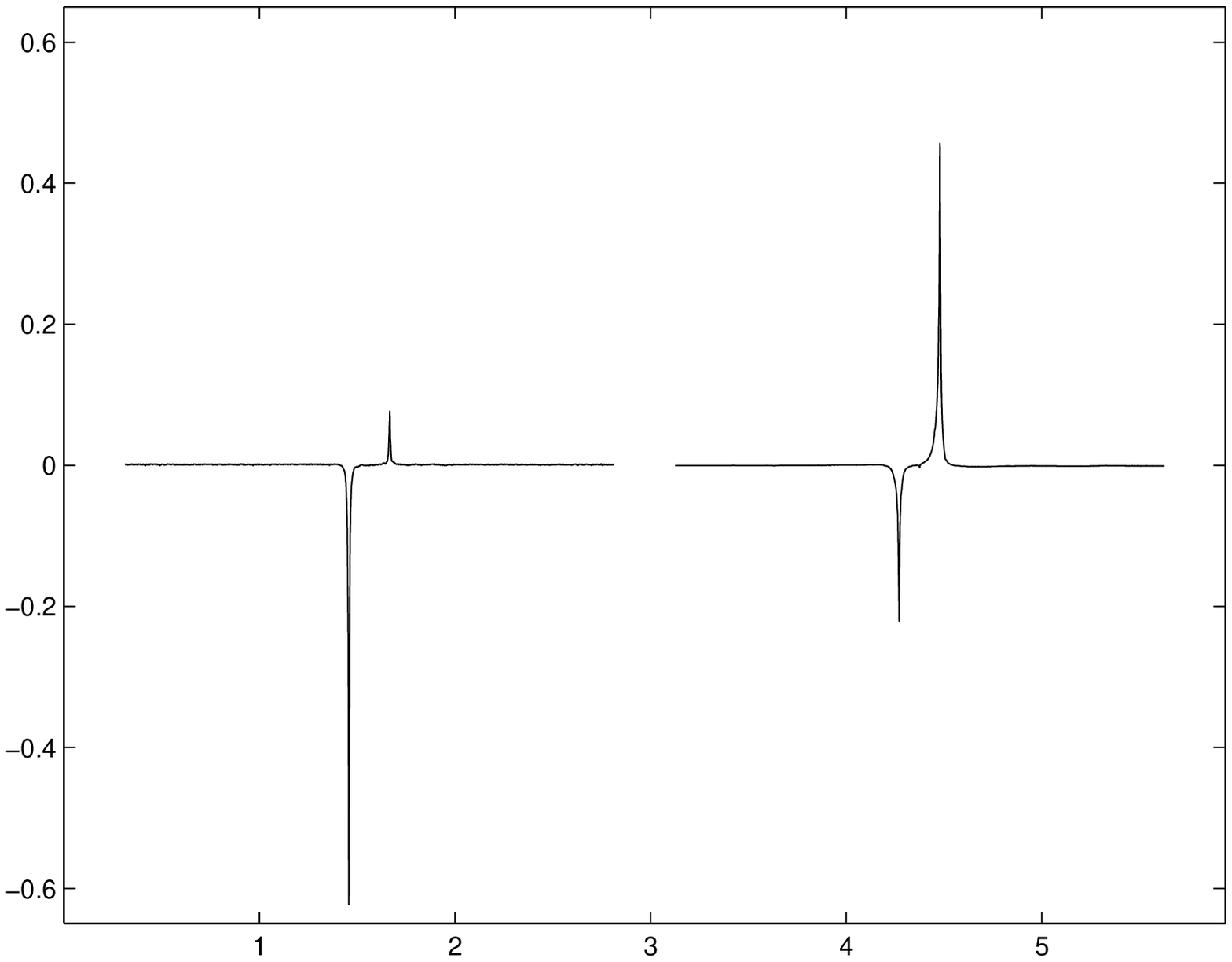,width=2.25in} }
\put(75,115){ $\LAB{B^CA^H}$ }
\put(170,5){ \psfig{file=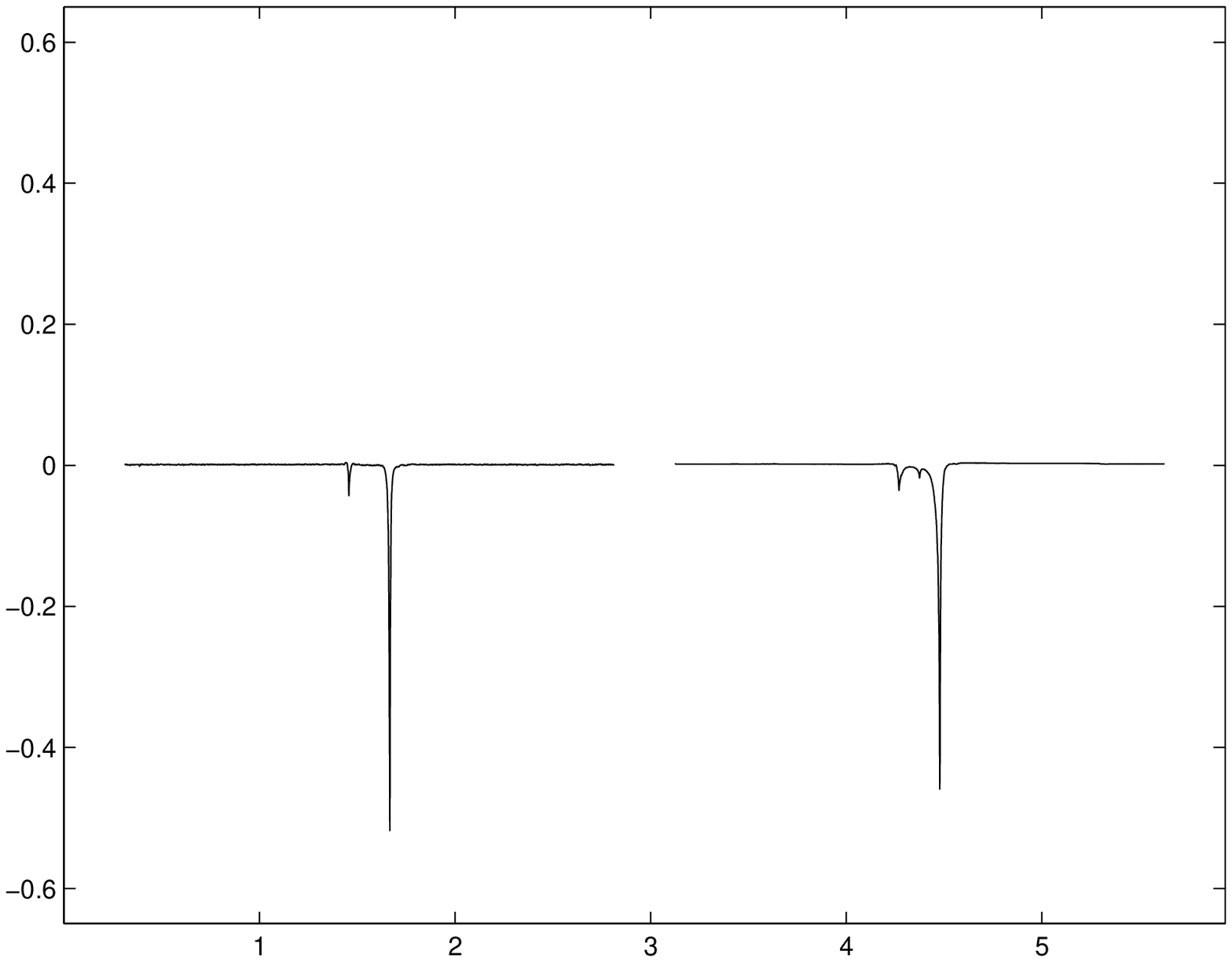,width=2.25in} }
\put(245,115){ $\LAB{B^CB^H}$ }
\end{picture} \caption{
The pairs of ${}^{13}\LAB{C}$-labeled chloroform
($\LAB{CHC}\ell_3$) spectra (carbon left, proton right)
obtained by performing the four combinations of measurements
$\LAB{A^CA^H}$, $\LAB{A^CB^H}$, $\LAB{B^CA^H}$ and $\LAB{B^CB^H}$
on the pseudo-pure form of Mermin's state $\EMB\Psi$.
The spectra have been normalized by the height of
the peak of the corresponding spin in the pseudo-pure
ground state, and the horizontal axis is in kHz.
The transitions of the peaks, from left to right, are
$\KET{0^\LAB{C}0^\LAB{H}} \leftrightarrow \KET{1^\LAB{C}0^\LAB{H}}$,
$\KET{0^\LAB{C}1^\LAB{H}} \leftrightarrow \KET{1^\LAB{C}1^\LAB{H}}$,
$\KET{0^\LAB{C}0^\LAB{H}} \leftrightarrow \KET{0^\LAB{C}1^\LAB{H}}$,
$\KET{1^\LAB{C}0^\LAB{H}} \leftrightarrow \KET{1^\LAB{C}1^\LAB{H}}$.}
\label{fig:hardy}
\end{figure}
\begin{table} \begin{center}
\renewcommand{\arraystretch}{1.1}%
\begin{tabular}{|c|rrrr|c|} \hline
\multicolumn{6}{|c|}{\parbox[t]{3.625in}{\bf
\centerline{
Probabilities of Outcomes {\boldmath$\LAB{G}$}
\& {\boldmath$\LAB{R}$} for the Measurements
}\centerline{
{\boldmath$\LAB{A}$} \& {\boldmath$\LAB{B}$}
(Carbon, Proton) Demonstrating Hardy's Paradox,
}\centerline{
as Derived from the Chloroform NMR Spectra
in Fig.\ \ref{fig:hardy}} \vspace{3pt}}
} \\ \hline\hline
~Measurements~ & $\quad(\LAB{G},\LAB{G})$ & $\qquad(\LAB{G},\LAB{R})$ &
$\qquad(\LAB{R},\LAB{G})$ & $\qquad(\LAB{R},\LAB{R})\quad$ & ~Residuals~
\\ \hline
$(\LAB{A^C},\LAB{A^H})$ & $0.253$ & $0.380$ & $0.366$ & $0.001\quad$
& $0.008$  \\
$(\LAB{A^C},\LAB{B^H})$ & $0.029$ & $0.609$ & $0.217$ & $0.145\quad$
& $0.018$  \\
$(\LAB{B^C},\LAB{A^H})$ & $-0.002$ & $0.230$ & $0.614$ & $0.159\quad$
& $0.005$ \\
$(\LAB{B^C},\LAB{B^H})$ & $0.097$ & $0.125$ & $0.156$ & $0.622\quad$
& $0.021$ \\ \hline
\end{tabular}
\renewcommand{\arraystretch}{1.3}%
\end{center} \end{table}

At this point we encounter a significant complication,
which is that the ``strong'' (von Neumann) measurements
assumed in their analyses by Jordan and
Mermin {\em cannot\/} be implemented by NMR;
we can only perform ``weak'' (nonperturbing) measurements
of the {\em probability differences\/} between states
connected by single spin flips \cite{CorPriHav:98}.
This is done by applying a magnetic field gradient
along the $\LAB{z}$-axis, which (as previously described)
dephases any transverse components in the density operator.
Thereafter, a pair of ``soft'' $\pi/2$ readout pulses,
each tuned to the frequency of just one of the two spins,
produces a pair of spectra each with two peaks whose
heights are proportional to the probability differences
between pairs of states connected by flips of that spin.
The factor relating the peak heights to the
corresponding differences in the probabilities
of the states can be determined from spectra
collected on the pseudo-pure ground state,
after which it is straightforward to convert
the differences into the corresponding
absolute probabilities by linear least squares,
subject to the constraint that their sum is unity.
We shall encounter field gradients again in the next section,
when we show how they can also be used to implement
precisely controlled decoherence models.

Thus the overall experiment consists of
collecting ten spectra, as follows:
\begin{enumerate}
\item Prepare the state $\EMB\Psi$, by first
preparing the pseudo-pure ground state $\KET{00}$
using one of the previously described methods,
and then transforming it by $\VEC Q(\theta) \VEC P(\phi)$.
\item Use a selective radio-frequency pulse to
apply the rotation $\VEC R$ to those spins on
which measurement $\LAB{B}$ is to be performed.
\item Use a $\LAB{z}$-gradient to dephase the transverse
components of the resulting density operator.
\item Apply a readout pulse to one of the
spins, and collect the corresponding spectrum;
repeat steps 1 -- 3 and then do the same for the other spin.
\item Repeat steps 1 - 4 for each of the four combinations of
measurements $\LAB{AA}$, $\LAB{AB}$, $\LAB{BA}$ and $\LAB{BB}$
on the two spins.
\item Collect two additional amplitude calibration spectra by applying
soft readout pulses to each spin in the pseudo-pure ground state.
\end{enumerate}
These experiments were performed on a Bruker
400 MHz spectrometer using the two spin $\HALF$
nuclei in ${}^{13}\LAB{C}$-labeled chloroform.

The spectra obtained from steps 1 -- 5 of this
experiment are shown in Fig.~\ref{fig:hardy}.
The probabilities derived from these peak heights,
and the residual (square-root of the sum of squares of the
deviations of the data points from the corresponding fit)
associated with each, are shown in the table below.
It follows that Bell's inequality is violated by
\begin{equation} \begin{array}{rl}
& \Psi_\LAB{GG}^\LAB{AB} + \Psi_\LAB{RR}^\LAB{AA} +
\Psi_\LAB{GG}^\LAB{BA} - \Psi_\LAB{GG}^\LAB{BB} \\
=~ & 0.029 + 0.001 - 0.002 - 0.097 ~=~ -0.069 ~.
\end{array} \end{equation}
A rigorous error analysis is not possible,
because the dominant errors in NMR spectra
(e.g.~RF field inhomogeneity) are not statistical.
If we nevertheless take the mean RMS residual
(half the total residuals shown in the table)
of $0.0065$ as an estimate of the errors and
assume they are independent between spectra,
the expected error in the sum of these four
numbers is only $0.013$, so that this violation
of Bell's inequality appears significant.

Nonetheless, as stressed in a recent preprint by Braunstein,
Caves, Jozsa, Linden, Popescu and Schack \cite{BraunsteinEtAl:98},
such experiments on weakly polarized pseudo-pure states
cannot actually disprove the existence of ``hidden variables''
associated with the spins of the individual molecules.
This is because the vast majority of a weakly polarized
density operator is contained in its identity component,
and there are many different ensembles of uncorrelated
spin states whose net density operator is the identity.
Hence the noise from the identity component dominates
the statistics of observations on the ensemble,
which are therefore consistent with microscopic
interpretations in which only uncorrelated
states are present with nonzero probability.
Indeed, if one were to pull the molecules
out the pseudo-pure sample used in the above
experiments one at a time, break them apart,
and perform the measurements $\LAB{A}$ and
$\LAB{B}$ with a Stern-Gerlach apparatus,
the frequencies of the four combinations of
outcomes would all be very close to $1/4$,
and would {\em not\/} violate Bell's inequality.
Thus, our apparent violation vanishes when the
whole ensemble is taken into consideration.

To see more precisely why the above experiments
fail to disprove the existence of hidden variables,
we first note that a pure state $\KET{\xi}\BRA{\xi}$ is
canonically associated with any given pseudo-pure density
operator $\RHO = (1-\delta)/2^N + \delta\,\KET{\xi}\BRA{\xi}$,
which is distinguished mathematically by the fact
that $\KET{\xi}$ is the eigenvector corresponding
to its {\em sole\/} nondegenerate eigenvalue.
We further recall (see Eq.\ (\ref{eq:pp_tfm})) that the
traceless part of the pseudo-pure density operator $\RHO$
transforms identically to that of the corresponding pure
state $\KET{\xi}\BRA{\xi}$ under unitary operations,
while the identity component transforms trivially,
and also that $\RHO$ produces exactly the same NMR spectrum
as would $\KET{\xi}\BRA{\xi}$ up to its overall amplitude
(since the identity component of any density operator
does not contribute to the signals observed by NMR).
Thus the unitary dynamics of the observables
in NMR experiments on pseudo-pure states are,
for all practical intents and purposes,
indistinguishable from the same experiments on a (smaller)
ensemble in the corresponding pure state $\KET{\xi}\BRA{\xi}$.

It follows that NMR experiments on pseudo-pure
states are necessarily {\em consistent with\/}
(though not proof of the reality of)
a microscopic interpretation of the ensemble in
which those molecules contributing to the observations
are all in the same pure state $\KET{\xi}\BRA{\xi}$,
while the remaining (and large majority of the)
molecules are in completely random states
with a net density operator of $1/2^N$.
In deriving a violation of Bell's inequality from
our measurements above, we required that the fractions
of molecules in the four diagonal states sum to unity,
so that they could be identified with the
probabilities of those states in an unidentified
\underline{sub\hspace*{0.1pt}}ensemble
in the pure state $\KET{\psi}\BRA{\psi}$.
Implicitly, therefore, this microscopic interpretation
of the ensemble was assumed in deriving the violation.
As explained above, however, the large identity
component in the corresponding pseudo-pure density
operator guarantees that many other ensembles
could be found with the same net density operator,
so that a microscopic interpretation in terms of
a single well-defined subensemble in the pure state
$\KET{\psi}\BRA{\psi}$ is not physically justified.
In fact, the fundamental limits on the amount of information
that can be extracted on an unknown quantum state even by strong
measurements prevents us from ever knowing if any molecules
of our pseudo-pure sample exist in or near the corresponding
pure state $\KET{\psi}\BRA{\psi}$ at all \cite{Peres:95}.
It is for this reason that our apparent violation of Bell's
inequality fails to disprove the existence of hidden variables.

This ambiguity in the microscopic interpretation
of liquid-state NMR experiments not-with-standing,
quantum physics indicates that a psuedo-pure spin state,
subjected to the same electromagnetic fields as a true
pure state, will undergo the same unitary transformation.
In addition, applying a $\LAB{z}$-gradient to an NMR ensemble
renders unobservable the same transverse phase information
that would be destroyed on performing strong measurements
along the $\LAB{z}$-axis on all the spins in the ensemble.
Finally, existing experiments relying upon true pure states
and strong measurements provide direct evidence against hidden
variable theories (see e.g.\ Ref.\ \cite{AspDalRog:82,Branning:97}).
Given this background knowledge of the underlying physics,
our experiments indirectly imply that the pure state
$\KET{\psi}\BRA{\psi}$ would violate Bell's inequality.
More generally, the ambiguity in the microscopic
interpretation of liquid-state NMR experiments in
{\em no way\/} detracts from their utility as a means
of studying the dynamics of information contained in
either pseudo-pure or (by inference) true pure states,
even in significantly more complex spin systems
that would be difficult to study by other means.
To further emphasize this fact, we will now describe
NMR experiments we have performed which demonstrate
quantum error correction using pseudo-pure states.

\vspace*{0.20in}
\section{Quantum error correction by NMR spectroscopy}
The error correcting code we have chosen to illustrate
by NMR is well-known in the field \cite{KnillLafla:97},
and uses two ancilla (labeled $2$ \& $3$) to
encode the state of a data spin (labeled $1$).
Letting $\VEC{S}^{2|1}$ and $\VEC{S}^{3|1}$ be
c-NOT's, and $\VEC{R}_{90}^{123} \equiv
\exp(-\imath\FRAC{\pi}{2}(\IY^1+\IY^2+\IY^3))$,
the encoding operation proceeds as follows:
\begin{equation} \begin{array}{rcl}
&& (\alpha\KET{0} + \beta\KET{1}) \KET{00} / \sqrt2 ~
\stackrel{\VEC{S}^{2|1}}{\longrightarrow}
\stackrel{\VEC{S}^{3|1}}{\longrightarrow}
\stackrel{\VEC{R}_{90}^{123}}{\longrightarrow}
~ \alpha \KET{\mbox{$+$$+$$+$}} + \beta\KET{\mbox{$-$$-$$-$}}
\\ &&
\left( \mbox{where}~ \KET{\mbox{$\pm$$\pm$$\pm$}}
\equiv (\KET{0}\pm\KET{1})(\KET{0}\pm\KET{1})
(\KET{0}\pm\KET{1}) / \sqrt8 \right)
\end{array} \end{equation}
Decoding consists of applying the
inverse operations in the reverse order,
which acts on the states obtained by
single sign-flip errors as follows:
\begin{equation} \begin{array}{rcl}
&&
\alpha \KET{\mbox{$+$$+$$-$}} + \beta\KET{\mbox{$-$$-$$+$}}
\stackrel{\VEC{R}_{-90}^{123}}{\longrightarrow}
\stackrel{\VEC{S}^{3|1}}{\longrightarrow}
\stackrel{\VEC{S}^{2|1}}{\longrightarrow}
(\alpha\KET{0} + \beta\KET{1})\KET{01} / \sqrt2
\\ &&
\alpha \KET{\mbox{$+$$-$$+$}} + \beta\KET{\mbox{$-$$+$$-$}}
\stackrel{\VEC{R}_{-90}^{123}}{\longrightarrow}
\stackrel{\VEC{S}^{3|1}}{\longrightarrow}
\stackrel{\VEC{S}^{2|1}}{\longrightarrow}
(\alpha\KET{0} + \beta\KET{1})\KET{10} / \sqrt2
\\ &&
\alpha \KET{\mbox{$-$$+$$+$}} + \beta\KET{\mbox{$+$$-$$-$}}
\stackrel{\VEC{R}_{-90}^{123}}{\longrightarrow}
\stackrel{\VEC{S}^{3|1}}{\longrightarrow}
\stackrel{\VEC{S}^{2|1}}{\longrightarrow}
(\alpha\KET{1} + \beta\KET{0})\KET{11} / \sqrt2
\end{array} \end{equation}
It follows that a Toffoli gate $\VEC{T}^{1|23}$,
which flips the data spin conditional on
the ancillae being in the state $\KET{11}$,
will correct a sign-flip error in the
data spin and leave it alone otherwise,
even if an error occurs in the ancillae.

In practice, errors in quantum computers
are not expected to be single sign-flips,
but rather small random phase errors
which cumulatively result in decoherence.
Nevertheless, we can show that the ability to
correct sign-flips implies the ability to cancel
the effect of such phase errors to first order.
Random phase errors correspond to the propagator
$\exp(-\imath(\chi^1\IZ^1+\chi^2\IZ^2+\chi^3\IZ^3))$,
where $\chi^1,\chi^2,\chi^3$ are random variables,
which acts to first order on the encoded state as:
\begin{equation} \begin{array}{rcl} &&
\exp(-\imath(\chi^1\IZ^1+\chi^2\IZ^2+\chi^3\IZ^3) \left(
\alpha \KET{\mbox{$+$$+$$+$}} + \beta\KET{\mbox{$-$$-$$-$}}
\right) \\ &~\approx~& \left(
\alpha \KET{\mbox{$+$$+$$+$}} + \beta\KET{\mbox{$-$$-$$-$}}
\right) - \imath\chi^1 \left(
\alpha \KET{\mbox{$-$$+$$+$}} + \beta\KET{\mbox{$+$$-$$-$}}
\right) \\ && -\, \imath\chi^2 \left(
\alpha \KET{\mbox{$+$$-$$+$}} + \beta\KET{\mbox{$-$$+$$-$}}
\right) - \imath\chi^3 \left(
\alpha \KET{\mbox{$+$$+$$-$}} + \beta\KET{\mbox{$-$$-$$+$}}
\right)
\end{array} \end{equation}
Since decoding and the error-correcting
Toffoli gate are likewise linear,
it follows that the first-order effects
of phase errors are cancelled as claimed.
Note this argument makes no assumptions
concerning the correlations among the errors!

Experimental results demonstrating these expectations
have recently been published \cite{CMPKLZHS:98}.
In the following, we shall present a more
detailed explanation of how the error correction
works using the product operator formalism,
along with selected experimental data
illustrating and validating this explanation.
We shall assume that the data spin is in one of the
states $1$ (unpolarized), $\IX^1$, $\IY^1$ or $\IZ^1$.
Although these are mixed states,
each consists of an incoherent sum of pure states,
e.g.\ $2\IZ^1 = \KET{0}\BRA{0} - \KET{1}\BRA{1}$,
so if error correction works on these pure states,
by linearity it will also work on the mixtures (and vice versa).
In these terms, a complete set of initial states
${\RHO}_\LAB{A}$ for error correction are:
\begin{equation} \begin{array}{rcl} \label{eq:rhoA}
&& \left. \begin{array}{l}
\EP^2\EP^3 \\ \IX^1 \EP^2\EP^3 \\
\IY^1 \EP^2\EP^3 \\ \IZ^1 \EP^2\EP^3
\end{array} \right\}
~\equiv~ {\RHO}_\LAB{A}^1 \EP^2 \EP^3 ~=~ {\RHO}_\LAB{A}
\end{array} \end{equation}
The corresponding states ${\RHO}_\LAB{B}$
to which they are mapped by encoding are:
\begin{equation} \begin{array}{rcl} \label{eq:rhoB}
{\RHO}_\LAB{B} &\equiv& \left\{ \begin{array}{l}
\FRAC14 + \IX^1\IX^2 + \IX^1\IX^3 + \IX^2\IX^3 \\
\IZ^1\IY^2\IY^3 + \IY^1\IZ^2\IY^3 + \IY^1\IY^2\IZ^3 - \IZ^1\IZ^2\IZ^3 \\
\IZ^1\IZ^2\IY^3 + \IZ^1\IY^2\IZ^3 + \IY^1\IZ^2\IZ^3 - \IY^1\IY^2\IY^3 \\
\FRAC{1}{4} (\IX^1 + \IX^2 + \IX^3) + \IX^1\IX^2\IX^3
\end{array} \right.
\end{array} \end{equation}
%\pagebreak
We note the last three states in Eq.~(\ref{eq:rhoA})
can be prepared (with a 50\% loss of polarization)
from the average of twice Eq.~(\ref{eq:condpure3a}) with
\begin{equation} \begin{array}{rcl} \label{eq:condpure3b}
&& \FRAC{1}{16}\, \MAT{Diag}( 3, 1, 1, 1, -3, -1, -1, -1 )
\\ &~\leftrightarrow~&
\FRAC{1}{16} (\IZ^1 (3 + 2\IZ^2 + 2\IZ^3 + 4\IZ^2\IZ^3))
\\ &=&
\FRAC{1}{4} (\EP^1 - \EM^1) (\EP^2\EP^3 + \FRAC{1}{2}) ~.
\end{array} \end{equation}

\begin{figure}[t]
\begin{picture}(300,250)
\put(5,0){ \psfig{file=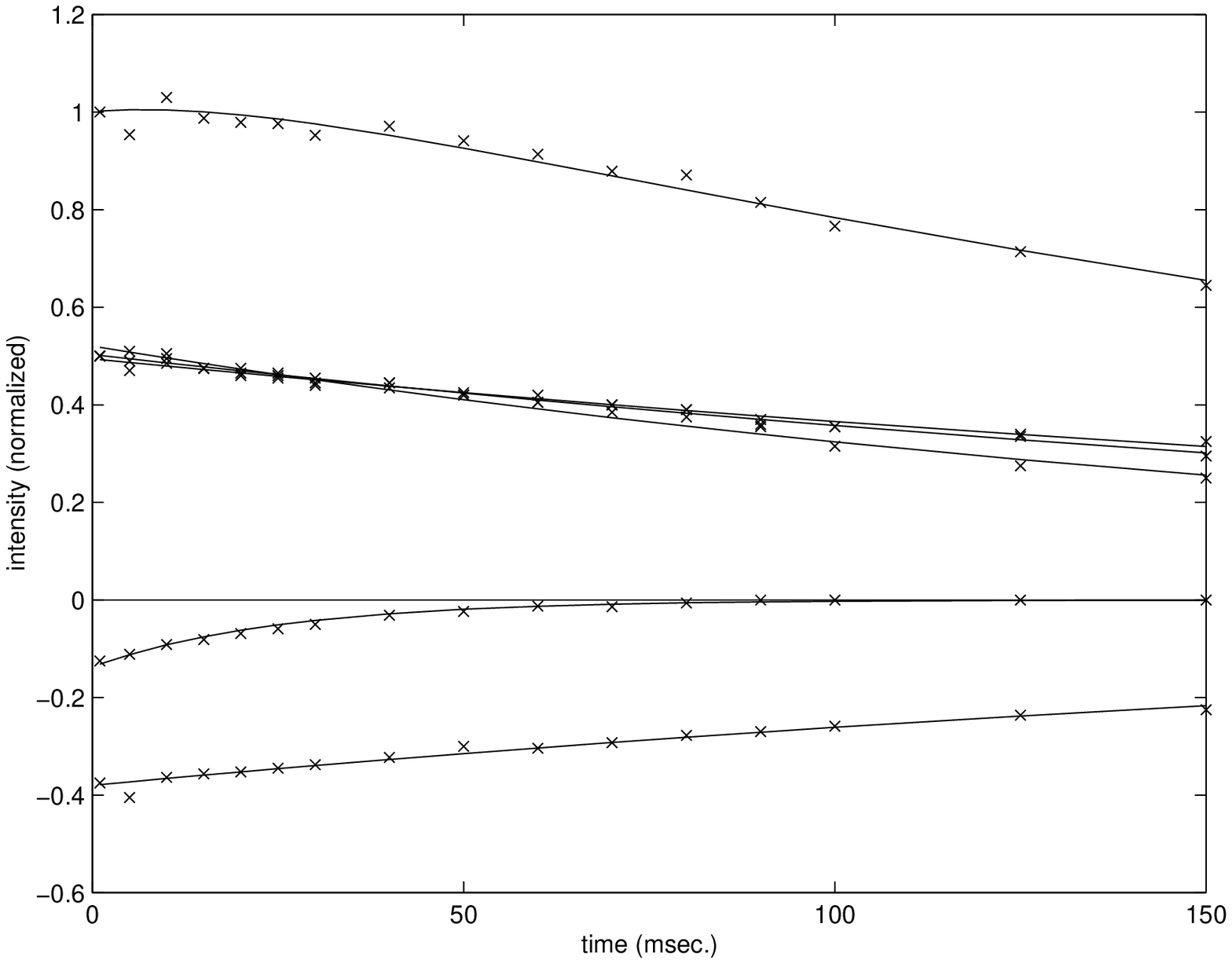,width=4.5in} }%,height=4.5in
\put(250,70){ $4\IZ^1\IZ^2\IZ^3$ (single) }
\put(250,100){ $4\IZ^1\IZ^2\IZ^3$ (triple) }
\put(250,145){ $\IZ^1$, $2\IZ^1\IZ^2$, $2\IZ^1\IZ^3$ }
\put(285,190){ TOTAL }
\end{picture} \caption{
Experimental NMR data illustrating the decay of each of the product
% EXPANDING THE FOLLOWING MACROS PRODUCES LINE TOO LONG IN AUX FILE
% operators $\IZ^1$, $2\IZ^1\IZ^2$, $2\IZ^1\IZ^3$ and $4\IZ^1\IZ^2\IZ^3$
operators $\mbox{\boldmath$I$}_{\sf z}^1$,
$2\mbox{\boldmath$I$}_{\sf z}^1\mbox{\boldmath$I$}_{\sf z}^2$,
$2\mbox{\boldmath$I$}_{\sf z}^1\mbox{\boldmath$I$}_{\sf z}^3$,
$4\mbox{\boldmath$I$}_{\sf z}^1\mbox{\boldmath$I$}_{\sf z}^2
\mbox{\boldmath$I$}_{\sf z}^3$,
as functions of the time allowed for gradient diffusion (see text),
together with the least-squares fits to their logarithms.
The single and triple quantum coherences in
% $4\IZ^1\IZ^2\IZ^3$	<-- DITTO!
$4\mbox{\boldmath$I$}_{\sf z}^1\mbox{\boldmath$I$}_{\sf z}^2
\mbox{\boldmath$I$}_{\sf z}^3$,
(negative curves) have been plotted and fit separately.
The sum of these data and of the fits are also shown (topmost curve),
which illustrates that error correction cancels the decay
of the encoded state to first order as expected.} \label{fig:err_cor}
\vspace*{-10pt}
\end{figure}

In liquid-state NMR, decoherence occurs principally
through the randomly fluctuating external magnetic fields
$B_\LAB{z}^k$ along the $\LAB{z}$-axis at each spin $k$.
The effect of these fields is most simply described
in the {\em spherical\/} product operator basis $\VEC 1$,
$\IZ^k$ and $\IPM^k \equiv \IX^k \pm \imath \IY^k$,
as opposed to the {\em Cartesian\/} basis used up to now.
The products of these basis elements can be shown \cite{ErnBodWok:87}
to decay exponentially at rates proportional to the
mean-square field $\OL{(B_\LAB{z}^k)^2}$ for $\IPM^k$
(as well as $\IPM^k \IZ^\ell$, $\IPM^k \IZ^\ell \IZ^m$),
and to
\begin{equation} \label{eq:fields}
\begin{array}[t]{rll}
& \OL{(B_\LAB{z}^k - B_\LAB{z}^\ell)^2} &
\quad\mbox{for $\IP^k\IM^\ell$ \& $\IM^k\IP^\ell$,} \\
& \OL{(B_\LAB{z}^k + B_\LAB{z}^\ell)^2} &
\quad\mbox{for $\IP^k\IP^\ell$ \& $\IM^k\IM^\ell$,} \\ 
& \OL{(B_\LAB{z}^k + B_\LAB{z}^\ell - B_\LAB{z}^m)^2} &
\quad\mbox{for $\IP^k\IP^\ell\IM^m$ \& $\IM^k\IM^\ell\IP^m$,~~etc.,} \\
\mbox{and}\quad & \OL{(B_\LAB{z}^k + B_\LAB{z}^\ell + B_\LAB{z}^m)^2} &
\quad\mbox{for $\IP^k\IP^\ell\IP^m$ \& $\IM^k\IM^\ell\IM^m$.}
\end{array}
\end{equation}
These products are referred to as single (SQC1: $\IPM^k$),
zero (ZQC: $\IPM^k\IMP^\ell$), double (DQC: $\IPM^k\IPM^\ell$),
three-spin single (SQC3: $\IPM^k\IPM^\ell\IMP^m$, etc.) and triple
(TQC: $\IPM^k\IPM^\ell\IPM^m$) quantum coherences, respectively.

We shall consider two extreme forms of decoherence.
In the first, the fields at the different spins are
uncorrelated, and hence the random variables $\chi^k$ can
be assumed to be identically distributed and independent.
In the second, they are assumed to be totally correlated.
By Eq.\ (\ref{eq:fields}), the relative rates
of decoherence in these two cases are:
\begin{equation}
\begin{array}{lccccc}
& ~\mbox{\sf ZQC}~ & ~\mbox{\sf SQC1}~
& ~\mbox{\sf SQC3}~ & ~\mbox{\sf DQC}~
& ~\mbox{\sf TQC}~ \\
\mbox{\sf Uncorrelated:}
& 2 & 1 & 3 & 2 & 3 \\
\mbox{\sf Totally Correlated:}
& 0 & 1 & 1 & 4 & 9
\end{array}
\end{equation}
Decomposing ${\RHO}_\LAB{B}$ into a spherical basis,
multiplying by decaying exponentials with the above
rates normalized by the SQC1 decay rate $\tau$,
and returning to the Cartesian basis gives
\begin{equation}
{\RHO}_\LAB{C} ~\equiv~ \left\{ \begin{array}{l}
\FRAC14 + (\IX^1\IX^2 + \IX^1\IX^3 + \IX^2\IX^3) e^{-2t/\tau}
\skiplinehalf \\
(\IZ^1\IY^2\IY^3 + \IY^1\IZ^2\IY^3 + \IY^1\IY^2\IZ^3) e^{-2t/\tau}
- \IZ^1\IZ^2\IZ^3
\skiplinehalf \\
(\IZ^1\IZ^2\IY^3 + \IZ^1\IY^2\IZ^3 + \IY^1\IZ^2\IZ^3) e^{-t/\tau}
- \IY^1\IY^2\IY^3 e^{-3t/\tau}
\skiplinehalf \\
\FRAC{1}{4} (\IX^1 + \IX^2 + \IX^3) e^{-t/\tau}
+ \IX^1\IX^2\IX^3 e^{-3t/\tau}
\end{array} \right.
\end{equation}
in the uncorrelated case, and
\begin{equation}
{\RHO}_\LAB{C} ~\equiv~
\left\{ \begin{array}{l}
\FRAC14 + \HALF (\IX^1\IX^2 + \IX^1\IX^3 + \IX^2\IX^3 +
\IY^1\IY^2 + \IY^1\IY^3 + \IY^2\IY^3) \\ +\,
\HALF (\IX^1\IX^2 + \IX^1\IX^3 + \IX^2\IX^3 - \IY^1\IY^2
- \IY^1\IY^3 - \IY^2\IY^3) e^{-4t/\tau}
\skiplinehalf \\
\HALF (\IZ^1(\IX^2\IX^3+\IY^2\IY^3) +
\IZ^2(\IX^1\IX^3+\IY^1\IY^3) +
\IZ^3(\IX^1\IX^2+\IY^1\IY^2)) \\ -\,
\HALF (\IZ^1(\IX^2\IX^3-\IY^2\IY^3) +
\IZ^2(\IX^1\IX^3-\IY^1\IY^3) +
\IZ^3(\IX^1\IX^2-\IY^1\IY^2)) \\ \qquad
e^{-4t/\tau} - \IZ^1\IZ^2\IZ^3
\skiplinehalf \\
(\IZ^1\IZ^2\IY^3 + \IZ^1\IY^2\IZ^3 + \IY^1\IZ^2\IZ^3)
e^{-t/\tau} \\ +\, \FRAC14
(\IX^1\IX^2\IY^3 + \IX^1\IY^2\IX^3 + \IY^1\IX^2\IX^3 + 3\IY^1\IY^2\IY^3)
e^{-t/\tau} \\ -\, \FRAC14
(\IX^1\IX^2\IY^3 + \IX^1\IY^2\IX^3 + \IY^1\IX^2\IX^3 - \IY^1\IY^2\IY^3)
e^{-9t/\tau}
\skiplinehalf \\
\FRAC14 (\IX^1 + \IX^2 + \IX^3)
e^{-t/\tau} \\ +\, \FRAC14
(3\IX^1\IX^2\IX^3 + \IY^1\IY^2\IX^3 + \IY^1\IX^2\IY^3 + \IX^1\IY^2\IY^3)
e^{-t/\tau} \\ +\, \FRAC14
(\IX^1\IX^2\IX^3 - \IY^1\IY^2\IX^3 - \IY^1\IX^2\IY^3 - \IX^1\IY^2\IY^3)
e^{-9t/\tau}
\end{array} \right.
\end{equation}
in the totally correlated case.
The decoding operation converts this to
\begin{equation}
{\RHO}_\LAB{D} ~\equiv~ \left\{ \begin{array}{l}
\FRAC14 + (\IZ^2 + \IZ^3 + \IZ^2\IZ^3) e^{-2t/\tau}
\skiplinehalf \\
(\HALF \IX^1\IZ^2 + \HALF \IX^1\IZ^3 + \IX^1\IZ^2\IZ^3) e^{-2t/\tau}
+ \FRAC{1}{4} \IX^1
\skiplinehalf \\
(\FRAC{1}{4} \IY^1 + \HALF \IY^1\IZ^2 + \HALF \IY^1\IZ^3) e^{-t/\tau}
+ \IY^1\IZ^2\IZ^3 e^{-3t/\tau}
\skiplinehalf \\
(\FRAC{1}{4} \IZ^1 + \HALF \IZ^1\IZ^2 + \HALF \IZ^1\IZ^3) e^{-t/\tau}
+ \IZ^1\IZ^2\IZ^3 e^{-3t/\tau}
\end{array} \right.
\end{equation}
in the uncorrelated case, and
\begin{equation}
{\RHO}_\LAB{D} ~\equiv~
\left\{ \begin{array}{l}
\FRAC14 + \HALF (\HALF\IZ^2 + \HALF\IZ^3 + \IZ^2\IZ^3 -
2\IX^1\IZ^2\IY^3 - 2\IX^1\IY^2\IZ^3 + \IY^2\IY^3) \\ +\,
\HALF (\HALF\IZ^2 + \HALF\IZ^3 + \IZ^2\IZ^3 +
2\IX^1\IZ^2\IY^3 + 2\IX^1\IY^2\IZ^3 - \IY^2\IY^3)
e^{-4t/\tau} 
\skiplinehalf \\
\HALF (\IX^1(\IY^2\IY^3+\IZ^2\IZ^3) +
\HALF \IZ^3(\IX^1-\IX^2) +
\HALF \IZ^2(\IX^1-\IX^3))
+ \FRAC{1}{4} \IX^1 \\ +\,
\HALF (\IX^1(\IY^2\IY^3-\IZ^2\IZ^3) +
\HALF \IZ^3(\IX^1+\IX^2) +
\HALF \IZ^2(\IX^1+\IX^3))
e^{-4t/\tau}
\skiplinehalf \\
(\HALF\IY^1\IZ^3 + \HALF\IY^1\IZ^2 + \FRAC{1}{4}\IY^1)
e^{-t/\tau} \\ +\, \FRAC14
(\IZ^1\IZ^2\IY^3 + \IZ^1\IY^2\IZ^3 - \IY^1\IY^2\IY^3 - 3\IY^1\IZ^2\IZ^3)
e^{-t/\tau} \\ -\, \FRAC14
(\IZ^1\IZ^2\IY^3 + \IZ^1\IY^2\IZ^3 - \IY^1\IY^2\IY^3 + \IY^1\IZ^2\IZ^3)
e^{-9t/\tau}
\skiplinehalf \\
\FRAC14 (\IZ^1 + 2 \IZ^1\IZ^2 + 2 \IZ^1\IZ^3)
e^{-t/\tau} \\ +\, \FRAC14
(3\IZ^1\IZ^2\IZ^3 + \IY^1\IY^2\IZ^3 + \IY^1\IZ^2\IY^3 + \IZ^1\IY^2\IY^3)
e^{-t/\tau} \\ +\, \FRAC14
(\IZ^1\IZ^2\IZ^3 - \IY^1\IY^2\IZ^3 - \IY^1\IZ^2\IY^3 - \IZ^1\IY^2\IY^3)
e^{-9t/\tau}
\end{array} \right.
\end{equation}
in the totally correlated case.
This is clearly getting a little messy,
and it gets much worse after the Toffoli gate!
Therefore, we shall only present the partial trace
over the ancillae after applying the Toffoli, which is
\begin{equation}
{\RHO}_\LAB{E}^1 ~\equiv~ \left\{ \begin{array}{l}
1 \\
\IX^1 \\
\IY^1 \left( \FRAC{3}{2} e^{-t/\tau} - \HALF e^{-3t/\tau} \right)
\\
\IZ^1 \left( \FRAC{3}{2} e^{-t/\tau} - \HALF e^{-3t/\tau} \right)
\end{array} \right.
\end{equation}
in the uncorrelated case, and
\begin{equation}
\RHO_\LAB{E}^1 ~\equiv~
\left\{ \begin{array}{l}
1 \\
\IX^1 \\
\IY^1 \left( \FRAC{3}{2} e^{-t/\tau} - \FRAC{3}{8} e^{-t/\tau}
- \FRAC{1}{8} e^{-9t/\tau} \right)
\\
\IZ^1 \left( \FRAC{3}{2} e^{-t/\tau} - \FRAC{3}{8} e^{-t/\tau}
- \FRAC{1}{8} e^{-9t/\tau} \right)
\end{array} \right.
\end{equation}
in the correlated.
The slope of these curves at $t = 0$ is zero in all cases, as expected.

In order to demonstrate these results by NMR solution-state spectroscopy,
a precise implementation of the above decoherence models is needed.
This was achieved by combining {\em gradient\/}
methods with molecular diffusion.
In these methods, a magnetic field gradient
is created along the $\LAB{z}$-axis;
as previously described, this dephases
the transverse ($\LAB{xy}$) magnetization.
More precisely, a field gradient causes the
transverse magnetization to precess at rates
which depend linearly on its $\LAB{z}$-coordinate,
thereby winding it into a spiral about the $\LAB{z}$-axis
whose average transverse magnetization is essentially zero.
The gradient is turned off for a given time interval $t$, during
which diffusion of the molecules along $\LAB{z}$ blurs the spiral.
The gradient is then reversed, causing the
magnetization to refocus and so create an ``echo''.
Because those molecules which have moved now precess at
a different rate, their magnetization is not refocussed,
so the magnitude of the echo decays exponentially with $t$.
Because all the spins in each molecule are subject
to the same change in field, this constitutes a
true implementation of the totally correlated model.
By using refocusing $\pi$-pulses between gradients,
it is also possible to dephase each spin separately,
thereby implementing the uncorrelated model.
At this time, however, we have collected and processed
data only for the ${\RHO}_\LAB{A}^1 = \IZ^1$
state with the totally correlated model.

Although it is possible to prepare the
state $\IZ^1\EP^2\EP^3$ as noted above,
we have chosen to illustrate the above analysis by
preparing the states $\IZ^1$, $2\IZ^1\IZ^2$, $2\IZ^1\IZ^3$
and $4\IZ^1\IZ^2\IZ^3$ in four separate experiments,
each using sixteen different decoherence times $t$.
Because the SQC and TQC contributing to $4\IZ^1\IZ^2\IZ^3$
refocussed at different times, this further
enabled us to follow their evolutions separately.
The results of these experiments are plotted
against the time $t$ in Figure \ref{fig:err_cor},
along with the corresponding logarithmic fits.
It may be seen that the sum of the data
and of the fits thereto (also shown) do
indeed exhibit a near-zero initial slope,
in accord with the above calculations.
Our published report \cite{CMPKLZHS:98}
includes the results of further experiments
(performed by E.~Knill and R.~Laflamme)
with the natural and far more complicated
decoherence processes that occur in solution.
These are more difficult to interpret,
but are nevertheless consistent with the state
preservation expected from error correction.
Additional experiments and more detailed
calculations are in progress.

While a method of inhibiting decoherence ($T_2$ relaxation)
during NMR pulse sequences would be highly desirable,
there are strong reasons to doubt that quantum
error correction will be useful in this regard.
First, the ancillae must be placed in a pseudo-pure state,
which as we have shown above entails a loss
of 50\% of the signal for each ``data'' spin;
this is more than is recovered by error correction.
In addition, the ancillae must be returned to a pseudo-pure
state uncorrelated with the state of the data spin(s), or else
``fresh'' ancillae in such a state must be continuously available,
in order to inhibit decoherence over an appreciable
period of time by the repeated correction of errors.
Nevertheless, we feel that the basic idea underlying
error correction of preparing multiple quantum coherences,
allowing them to decohere, and then mixing them so
as to determine their relative rates of relaxation,
may be of considerable use in NMR studies
of the statistics of molecular motion.
This in turn is one of the most important
applications of NMR spectroscopy.
Conversely, whereas NMR spectroscopists have
previously used their methods solely to unravel
the secrets of naturally occurring systems,
it now appears possible to use these same methods
to engineer artificial systems in which the basic
principles of quantum information processing,
in particular the emergence of the classical
world through decoherence \cite{GiuliniEtAl:96},
can be studied in unprecedented detail.

%\vspace{-10pt}
\section*{ACKNOWLEDGEMENTS}
%\begin{acknowledgement}
We thank E.~Knill and R.~Laflamme of
Los Alamos National Labs for teaching
us about quantum error correcting codes,
and S.~Braunstein and R.~Jozsa for useful
discussions on mixed-state correlation.
This work was supported by the U.~S.~Army Research
Office under grant number DAAG 55-97-1-0342
from the DARPA Ultrascale Computing Program.
%\end{acknowledgement}
%\pagebreak

\bibliography{../../math,../../csci,../../nmr,../../phys,../../self}
% ieeetr gets bibtex to produce output more or less as the AAECC wants,
\bibliographystyle{ieeetr}
\end{document}